\documentclass[utf8]{frontiersSCNS} % for Science, Engineering and Humanities and Social Sciences articles

\usepackage{url,hyperref,lineno,microtype,subcaption}
\usepackage[onehalfspacing]{setspace}
\usepackage{aas_macros} 
\usepackage{upgreek}
\newcommand{\kplr}{{\it Kepler}}
\newcommand{\cd}{d$^{-1}$}

\def\keyFont{\fontsize{8}{11}\helveticabold }
\def\firstAuthorLast{D.~L.~Holdsworth}
\def\Authors{Daniel L. Holdsworth\,$^{*}$}

\def\Address{Jeremiah Horrocks Institute, University of Central Lancashire, Preston PR1 2HE, UK}

\def\corrAuthor{Daniel L. Holdsworth}

\def\corrEmail{dlholdsworth@uclan.ac.uk}

\begin{document}
\onecolumn
\firstpage{1}

\title[roAp stars and Kepler]{The roAp stars observed by the {\it Kepler} Space Telescope} 

\author[\firstAuthorLast ]{\Authors} %This field will be automatically populated
\address{} %This field will be automatically populated
\correspondance{} %This field will be automatically populated

\extraAuth{}% If there are more than 1 corresponding author, comment this line and uncomment the next one.
%\extraAuth{corresponding Author2 \\ Laboratory X2, Institute X2, Department X2, Organization X2, Street X2, City X2 , State XX2 (only USA, Canada and Australia), Zip Code2, X2 Country X2, email2@uni2.edu}
\maketitle

\begin{abstract}
Before the launch of the {\it Kepler} Space Telescope, most studies of the rapidly oscillating Ap (roAp) stars were conducted with ground-based photometric $B$ observations, supplemented with high-resolution time-resolved spectroscopy and some space observations with the WIRE, MOST and BRITE satellites. These modes of observation often only provided information on a single star at a time, however, {\it Kepler} provided the opportunity to observe hundreds of thousands of stars simultaneously. Over the duration of the primary 4-yr {\it Kepler} mission, and its 4-yr reconfigured K2 mission, the telescope observed at least 14 new and known roAp stars. This paper provides a summary the results of these observations, including a first look at the entire data sets, and provides a look forward to NASA's {\it TESS} mission.
\tiny
 \keyFont{ \section{Keywords:} asterosismology, stars: chemically peculiar, stars: oscillations, techniques: photometric,  stars: variables} %All article types: you may provide up to 8 keywords; at least 5 are mandatory.
\end{abstract}

\section{Introduction}

The rapidly oscillating, chemically peculiar A (roAp) stars are found at the base of the classical instability strip, where it intersects the main-sequence. They provide the opportunity to study the interactions between strong magnetic fields, pulsation, rotation and chemical stratification. Due to these properties, the roAp stars provide a unique insight to stellar atmospheres in 3D.

The Ap stars as a class are characterised by spectroscopic signatures of Sr, Eu, Cr and/or Si in low-resolution classification spectroscopy \citep[e.g.,][]{2009ssc..book.....G}. In high-resolution spectroscopy, abundances of rare earth elements can be measured to be over a million times that observed in the Sun \citep[e.g.,][]{2010A&A...509A..71L}. They are permeated by a strong, global, magnetic field that can be up to about 35\,kG in strength \citep{2010MNRAS.402.1883E}. It is the presence of the magnetic field which gives rise to the chemical peculiarities in the Ap stars; convection is suppressed by the magnetic field allowing for the radiative levitation of, most notably the rare earth elements, and the gravitational settling of others. The origin of the magnetic field is not conclusively known, but is suspected to be the result of the merger of close binary stars in the pre-main sequence phase of evolution where at least one star is still on the Henyey track \citep{2009MNRAS.400L..71F,2010ARep...54..156T}. This scenario provides an explanation for the lack of Ap stars in close binary systems, and why the magnetic axis is inclined to the rotation axis.

The rapid oscillations in Ap stars were first discovered through the targeted observations of these stars by \citet{1978IBVS.1436....1K}, with the seminal paper following a few years later \citep{1982MNRAS.200..807K}. Since their discovery, over 75 roAp stars have been identified through ground- and space-based observations \citep[e.g.,][]{1994MNRAS.271..118M,2013MNRAS.431.2808K,2014MNRAS.439.2078H,2016A&A...590A.116J,2019MNRAS.488...18H,2019MNRAS.487.3523C}. They show pulsational variability in the range $5-24$\,min and pulsate in low degree ($\ell\leq3$), high-overtone ($n>15$) modes that are thought to be driven by the $\kappa$-mechanism acting on the H\,{\sc{i}} ionization zone \citep[e.g.,][]{2001MNRAS.323..362B,2005MNRAS.360.1022S}. However, this mechanism is unable to explain all of the observed pulsation frequencies in the roAp stars, leading \citet{2013MNRAS.436.1639C} to postulate that turbulent pressure may play a role in the excitation of some observed modes.

Given the $\kappa$-mechanism is the most likely driving mechanism for {\it most} of the pulsations in roAp stars the frequencies should be stable, unlike the stochastically excited modes that are seen in the solar-like and red giant stars. Some of the roAp stars show very stable pulsation modes over the entire data spans. Others, however, do not. The most well studied case of frequency variability in an roAp star is HR\,3831 where 16\,yr of ground-based data are available \citep{1994MNRAS.268..641K,1997MNRAS.287...69K}, with 8 other stars showing frequency variability identified by \citet{1994MNRAS.271..305M}. The cause of the frequency variability is unknown, but is postulated to have two possible origins. The first is frequency perturbation by an external body. In this instance, the Doppler shift of the roAp star is imparted on the arrival time, at the observer, of the pulsation signal. For a circular orbit, this would induce a sinusoidal change in the frequency. Much work has been done on this aspect in recent years, with the development of the Frequency Modulation (FM) theory \citep{2012MNRAS.422..738S,2015MNRAS.450.3999S}, with an extension to Phase Modulation \citep[PM;][]{2014MNRAS.441.2515M,2015MNRAS.450.4475M}. Since, when fitting a sinusoidal function to a pulsation mode, the frequency and phase are inextricably intertwined, so the discussion of frequency and phase modulation is one and the same. In the reminder of this work, frequency variability is discussed in the \kplr\ roAp stars, but plots showing phase changes are produced since this is the directly observed changing feature. 

The second interpretation of frequency variability in roAp stars is a change in the cavity in which the mode propagates. This could be due to an evolutionary change as the star evolves off the main-sequence. In this scenario, one would expect a monotonic decrease in the frequency \citep[as is seen in solar-like pulsators; e.g.,][]{2013ARA&A..51..353C}. Alternately, \citet{1994MNRAS.268..641K} suggested that cyclic variability could be an indication of a stellar magnetic cycle analogous to the 11-yr solar cycle. For non-cyclic and non-monotonic frequency variability small changes in the internal magnetic field configuration may be responsible. There could, of course, be a combination of many of these possibilities at work. With the ever increasing precision, and growing time base of observations, frequency variability is becoming a common observed feature of the roAp stars.

In fact, there is evidence for frequency variability in most pulsating stars. Stochastically excited modes in, for example, solar-like pulsators are incoherent which results in a natural variation in the pulsation frequency (or phase) over time. However, the classically pulsating stars are driven by a coherent force resulting in stable frequencies. Despite this, frequency variability is still observed. \citet{2016JAVSO..44..179N} discussed the observed frequency changes in a variety of classical pulsators in the context of evolutionary changes in the star. While some observations were accurately modelled by evolutionary changes alone, there was evidence for further physical processes altering the pulsation frequency. Whether these physical processes involve rotation, mass loss, magnetic fields, small changes in local chemical composition, or a combination of these factors is currently unclear. The identification and analysis of frequency changes in pulsating stars will pave the way for detailed stellar modelling to try and solve this problem.

The pulsations in roAp stars have overtones significantly greater than the degree, meaning the modes are in the asymptotic regime \citep{1979PASJ...31...87S,1980ApJS...43..469T,1990ApJ...358..313T}. In this case, p\,modes become regularly spaced in frequency with alternating odd and even degree modes \citep[see e.g.,][]{aerts2010}. The spacing between two consecutive modes of the same degree is the large frequency separation, $\Delta\nu$, while the frequency difference between two odd or even degree modes (e.g., $\ell=0$ and $\ell=2$ or $\ell=1$ and $\ell=3$) that have a radial overtone difference of one is the small frequency separation, $\delta\nu$. The value of the large frequency separation is dependent on stellar global properties, changing in proportion to the square root of the mean density, and can be scaled with reference to the Sun. The small frequency separation is sensitive to the core concentration of the star, and hence is an age indicator.

In the case of the roAp stars, the presence of a strong, global, magnetic field causes significant deviation from spherical symmetry and as such breaks the regular pattern of p\,modes. This `glitch' in the regular spacing changes $\Delta\nu$ for a single spacing, and may cause a change in the odd-even-odd pattern of the mode degree depending on the magnetic field geometry. The most well-studied case of this phenomenon is HR\,1217 \citep{2001MNRAS.325..373C,2005MNRAS.358..651K}.

From the early observations of roAp stars, it was seen that the pulsation mode in an amplitude spectrum is often accompanied by sidelobes that are split by the rotation frequency of the star. \citet{1982MNRAS.200..807K} interpreted this as oblique pulsation. It is well known that the magnetic field axis in the Ap stars is misaligned with the rotation axis, leading to the oblique rotator model \citep{1950MNRAS.110..395S}. This configuration results in a variable light curve over the rotation period of an Ap star, as chemical spots often form at the magnetic poles and provide a brightness contrast against the photosphere. The pulsation axis in a star is the axis of greatest deformation; in most stars it is the rotation axis that serves to break spherical symmetry, and in some binary stars it is the line of apsides \citep{2020NatAs.tmp...45H,2020MNRAS.494.5118K}. In the case of the Ap stars, it is the magnetic field, which can be of order 30\,kG \citep[e.g.,][]{1960ApJ...132..521B,2008MNRAS.389..441F,2017A&A...601A..14M}, that causes the most significant deviation from spherical symmetry, and so the pulsation axis is closely aligned to the magnetic one \citep[e.g.,][]{shibahashi1985,shibahashi1993,2002A&A...391..235B,2011A&A...536A..73B}. 

This misalignment of the pulsation axis with the rotation one serves to provide an observer with a varying view of the pulsation, leading to amplitude modulation of the observed pulsation(s). In a Fourier spectrum of a light curve, one expects to see a multiplet of $2\ell+1$ components for a {\it pure} mode, that is to say a mode that is described by a single spherical harmonic. For a {\it distorted} mode, the highest number of multiplet components so far observed is 14 (see the discussion of HD\,24355 below). To detect these multiplet components, observations are required to cover at least 1.5 rotation cycles of the star, with a greater number being preferable. With the presence of the sidelobes, it is possible, through the analysis of their amplitude ratios, to provide constraints on the geometry of the star. The stellar inclination value, $i$, and the angle between the rotation axis and the pulsation axis, $\beta$, can be either constrained or determined through the relations provided by \citet{1992MNRAS.259..701K}.

The rotation periods of the Ap stars are, on the whole, significantly longer than their non-magnetic, chemically normal, counterparts likely due to magnetic braking \citep{stepien2000}. It is common to find A stars with $v\sin i$ values greater than $120\,$km\,s$^{-1}$ \citep{2004IAUS..224....1A}, whereas the Ap stars are considerably lower \citep[up to $40\,$km\,s$^{-1}$;][]{1995ApJS...99..135A}. Indeed, the rotation periods for some Ap stars are thought to be as long as centuries \citep{2017A&A...601A..14M}. This serves as a problem when trying to apply the oblique pulsator model to determine the mode geometry. Although there is not yet a causal link established, over half of the known roAp stars have undetermined rotation periods, implying they are much longer than the observations cover \citep{2020arXiv200314144M}. It is hoped that the ongoing  Transiting Exoplanet Survey Satellite ({\it TESS}) mission \citep{2015JATIS...1a4003R} will provide insight to this.

\section{{\it Kepler} observations of roAp stars}

The \kplr\ Space Telescope \citep{2010ApJ...713L..79K} was launched in 2009 to a 372.5\,d Earth-trailing heliocentric orbit. The mission collected data in two cadences: the Long Cadence (LC) mode of 29.43\,min and the Short Cadence (SC) mode of 58.85\,s \citep{2010PASP..122..131G}.

The pulsation mode frequencies in roAp stars are significantly greater than the Nyquist frequency of the LC data ($24.469\,$\cd). This causes an attenuation of the mode amplitude due to the length of the integration -- the signal is effectively smeared out over the course of the 30-min observation. Furthermore, with exactly equally split data, Nyquist reflections of the true mode all have the same amplitude. However, \citet{2013MNRAS.430.2986M} showed that although the exposures triggered by the on-board clock are at set intervals, when the time stamps are corrected to Barycentric time, the regularity of the exposures is broken, allowing for the development of Super-Nyquist asteroseismology. As such, pulsations with frequencies above the LC Nyquist frequency can be observed, and distinguished from aliases. The optimal way of observing roAp stars with \kplr\ was with the SC mode, where the sampling is much greater than the pulsation period. However, for each month only 512 stars could be observed in SC mode, thus limiting the target selection for roAp stars.

\begin{figure}[htb]
    \centering
    \begin{minipage}{0.29\linewidth}
    \caption{HR diagram showing the location of the roAp stars where T$_{\rm eff}$ and luminosity estimates are available. The stars discussed here, which have been observed by \kplr, are shown by the red stars. The blue lines mark the extent of the theoretical instability strip as calculated by \citet{2002MNRAS.333...47C}. A selection of non-oscillating Ap stars are shown by dots. The evolutionary tracks, in solar masses, are from \citet{2008A&A...484..815B}. A representative error bar is shown in the lower left corner.}
    \label{fig:HRD}
    \end{minipage}
    \begin{minipage}{0.7\linewidth}
    \includegraphics[width=\linewidth]{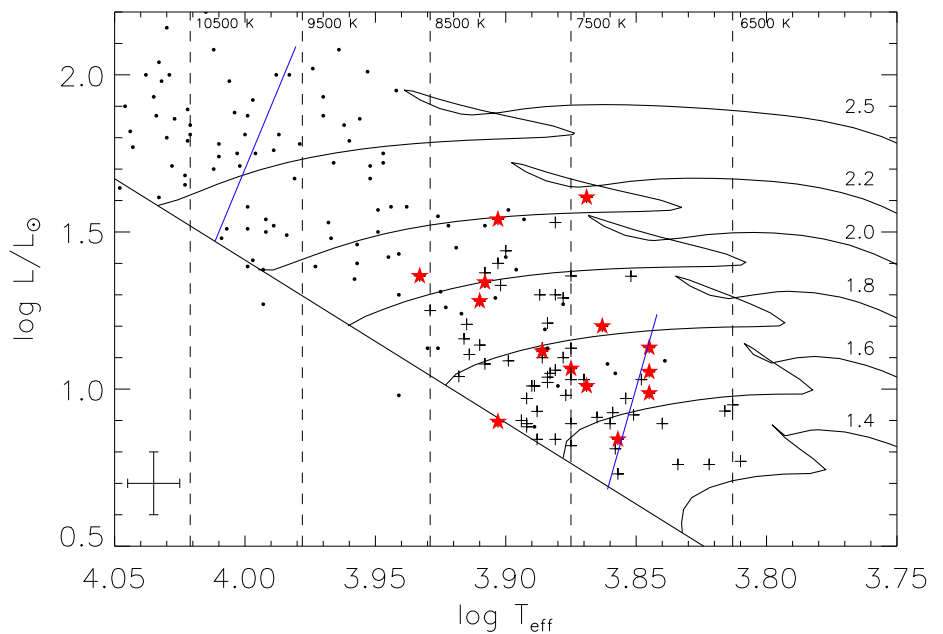}
    \end{minipage}
\end{figure}

The following sections will review the roAp stars observed during both the primary \kplr\ mission and the subsequent K2 mission. In total, \kplr\ observed 14 roAp stars in its primary and K2 configurations. These stars are plotted as red stars on a HR diagram in Fig.\,\ref{fig:HRD}, where they are shown alongside other roAp stars (black pluses) and Ap stars that do not show detectable pulsations (black dots). As demonstrated by this figure, the rapid oscillations are predominantly found in the cooler Ap stars, despite the theoretical instability strip (calculated under the assumption that the magnetic field suppresses convection in some region of the stellar envelope; \citealt{2002MNRAS.333...47C}) for these stars extending to about 10\,000\,K. This uneven distribution may be a result of the targeted ground-based observation campaigns from which many roAp stars were discovered \citep[e.g.,][]{1991MNRAS.250..666M,2012A&A...542A..89P,2016A&A...590A.116J}.\\

\subsection{Primary Mission}

At the time of the launch of \kplr\ there were no known roAp stars in the proposed field of view, and few known Ap stars. However, during the 4-yr primary mission, at least 11 roAp stars were observed. These stars were identified from the first 10-d commissioning run, up to using the full 4-yr data set to search for pulsations above the LC Nyquist frequency.  \\

\subsubsection{Long Cadence observations}

Seven roAp stars were observed by \kplr\ in LC mode only. The first, KIC\,7582608, was identified in SuperWASP photometry, with the \kplr\ data subsequently analysed by \citet{2014MNRAS.443.2049H}. The other 6 stars were identified as roAp stars by \citet{2019MNRAS.488...18H} via a search of the 4-yr data set in the super-Nyquist regime.

\begin{figure}
    \centering
    \includegraphics[width=0.5\textwidth]{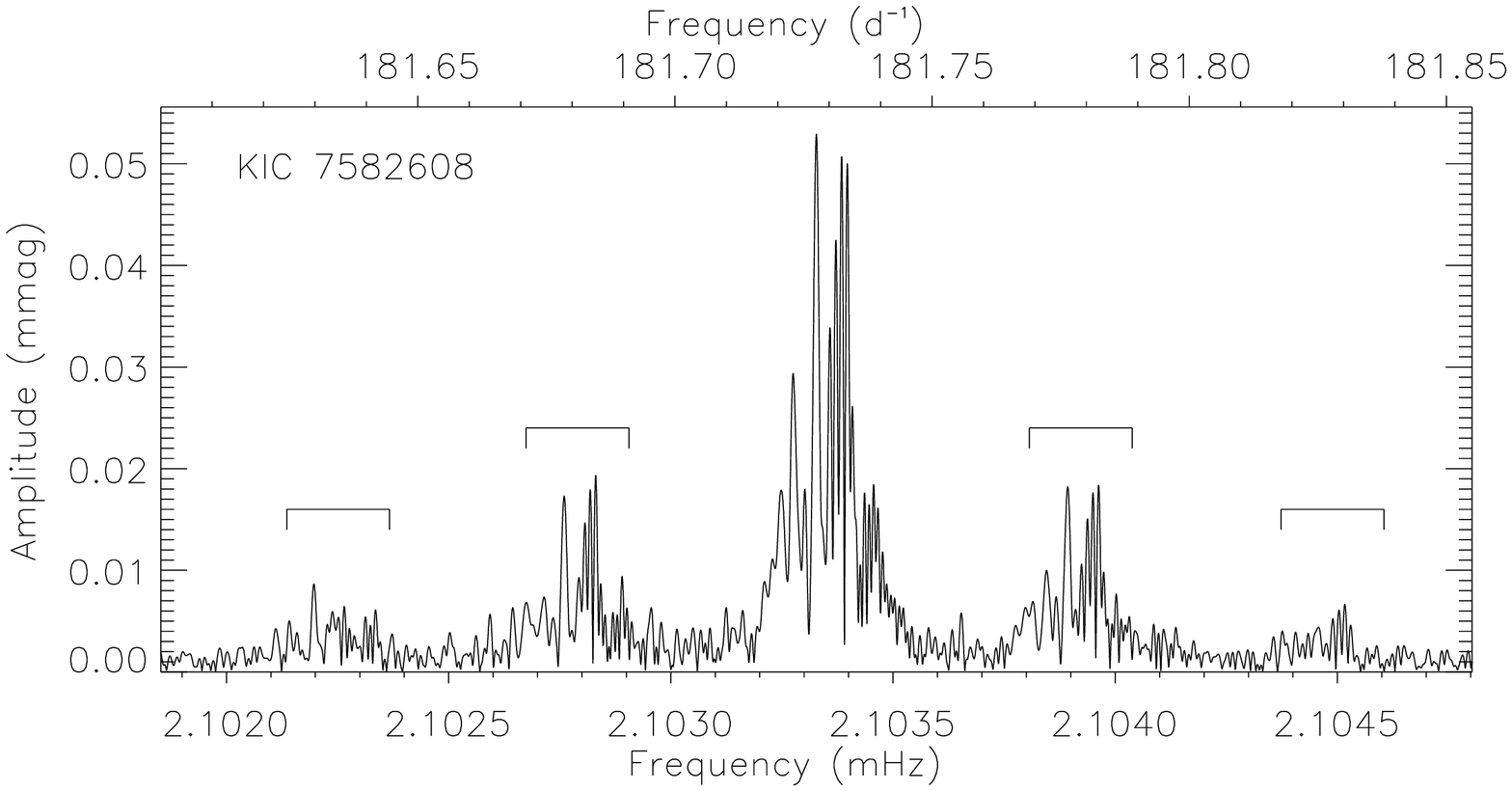}\hfill
    \includegraphics[width=0.5\textwidth]{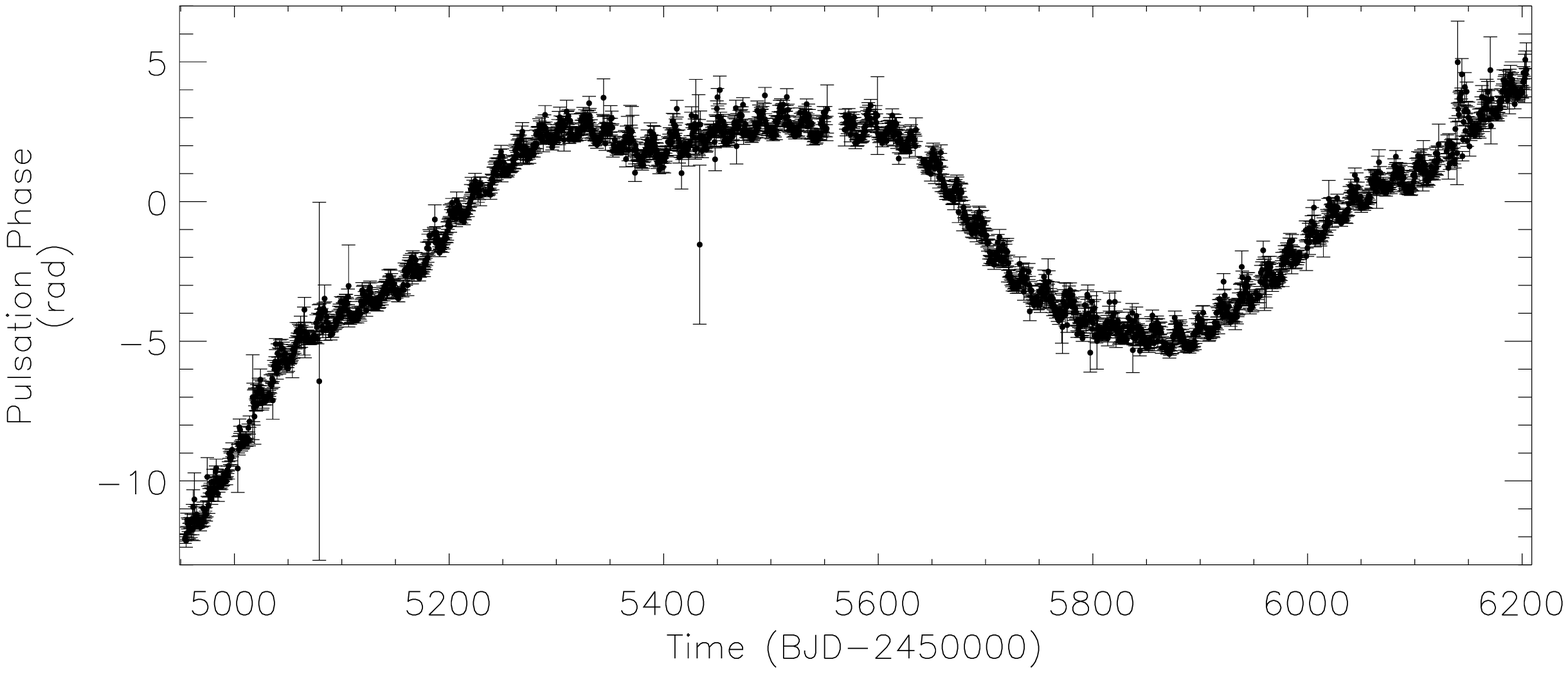}
    \caption{Left: amplitude spectrum of the LC 4-yr light curve of KIC\,7582608 showing the `ragged' structure discussed in the text. The five broad peaks, the central peak and those under the brackets,} are split by the rotation frequency of the star ($0.566\,\upmu$Hz, $P=20.4499$\,d), indicating a quadrupole mode. Right: phase variation of the pulsation mode indicating significant frequency variability in this star.
    \label{fig:7582608}
\end{figure}

The study of KIC\,7582608 in the super-Nyquist regime was aided by the knowledge of the pulsation frequency from ground-based observations. Although it is possible to identify the correct pulsation frequency from aliases, prior knowledge removes any ambiguity. The analysis of this star was further aided by its intrinsically large amplitude, such that the signal was not suppressed to a non-detectable level. The data for this star showed a single mode that was split into a quintuplet by oblique pulsation. However, the peaks of the quintuplet were not `clean' but described as `ragged', i.e. they are partially resolved into many closely spaced peaks. This is shown in Fig.\,\ref{fig:7582608}. Such an observation implies frequency and/or amplitude variability in a star such that a single frequency cannot describe the variability. This variability was investigated by fitting a fixed frequency to short sections of the data, and observing how the pulsation amplitude and phase changed. The amplitude was constant, after accounting for the variability caused by oblique pulsation, but the phase showed a significant change over the length of the observations.

There are two main interpretations of this observed phenomenon: intrinsic frequency variability caused by changes in the pulsation cavity, or an external body causing positional changes in the star which leads to frequency modulation (FM) in a pulsation mode \citep{2012MNRAS.422..738S}. Given the length of the data set, and the almost cyclic nature of the phase variability, \citet{2014MNRAS.443.2049H} proposed that binary motion was the cause of the changes observed in KIC\,7582608. By converting the pulsation frequency measured at discrete times into velocities, the authors showed a photometric RV curve. From these measurements, a binary model suggested a orbital period of about 1200\,d and a minimum mass of a companion of 1\,M$_\odot$. Subsequent spectroscopic RV measurements are currently inconclusive on the presence of a companion.

It was unfortunate that this star was not observed in SC mode at all during the \kplr\ mission, as high-precision, time-resolved observations may have provided the opportunity to resolve the binary/intrinsic variability conundrum in this star. For example, the identification of further, low-amplitude modes which behaved in the same way as the principal mode would suggest an external driving for the variability, otherwise a likely conclusion would be changes within the star which affect different modes differently.

The other six roAp stars observed with \kplr\ in LC mode were KIC\,6631188, KIC\,7018170, KIC\,10685175, KIC\,11031749, KIC\,11296437 and KIC\,11409673 (Fig.\,\ref{fig:HEY_stars}).
These stars were identified in a novel way by \citet{2019MNRAS.488...18H}. By selecting all stars in the \kplr\ data set with T$_{\rm eff}>6000$\,K, they identified variable stars by calculating the skewness of high-pass filtered light curves, and searching for non-aliased peaks. This technique is sensitive to most pulsation frequencies apart from those close to integer multiples of the sampling frequency, and the very highest frequencies (above about 3500\,$\upmu$Hz; 300\,\cd) in the \kplr\ LC data. 

\begin{figure}
    %\centering
    \includegraphics[width=0.5\textwidth]{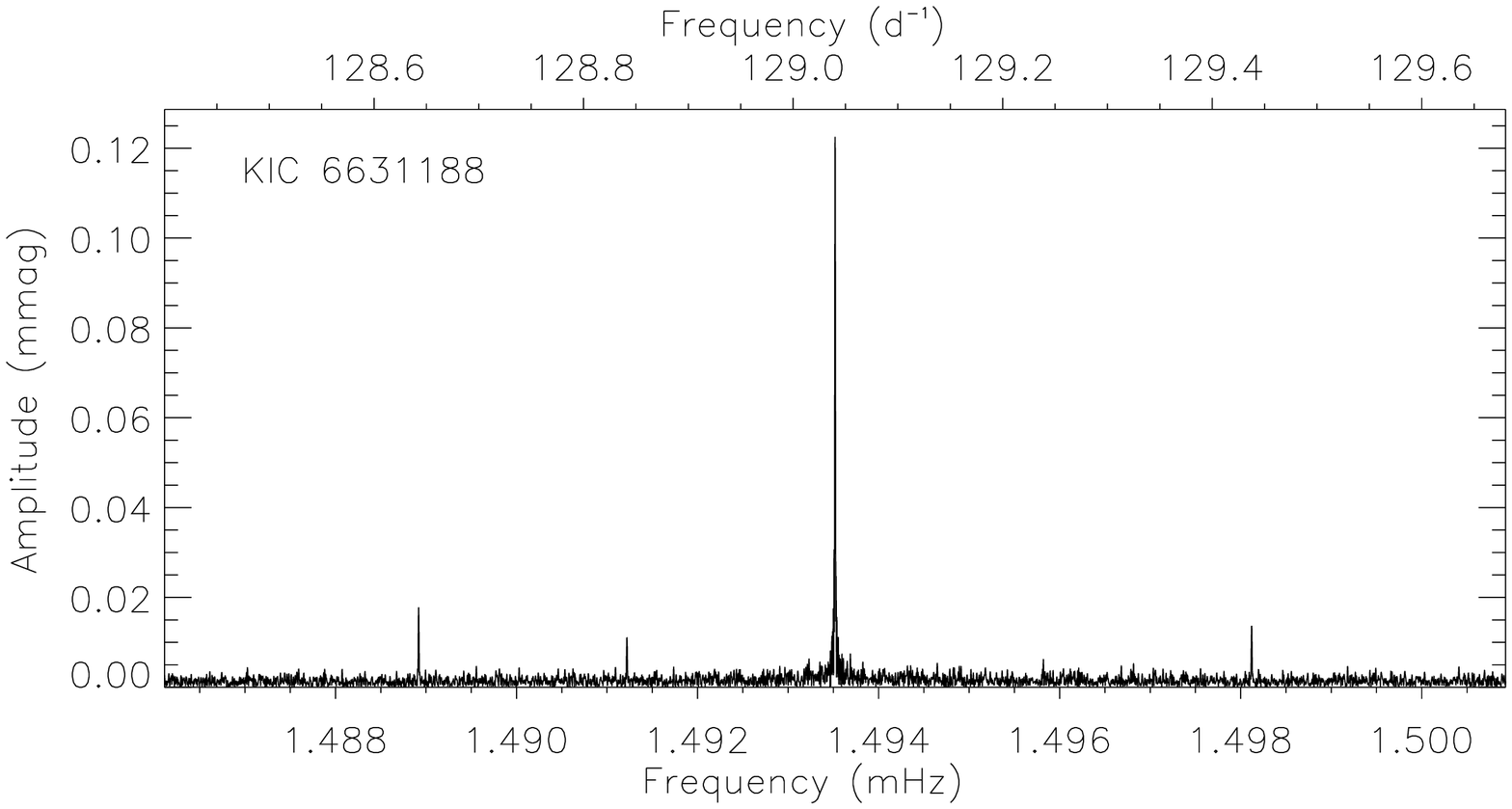}
    \includegraphics[width=0.5\textwidth]{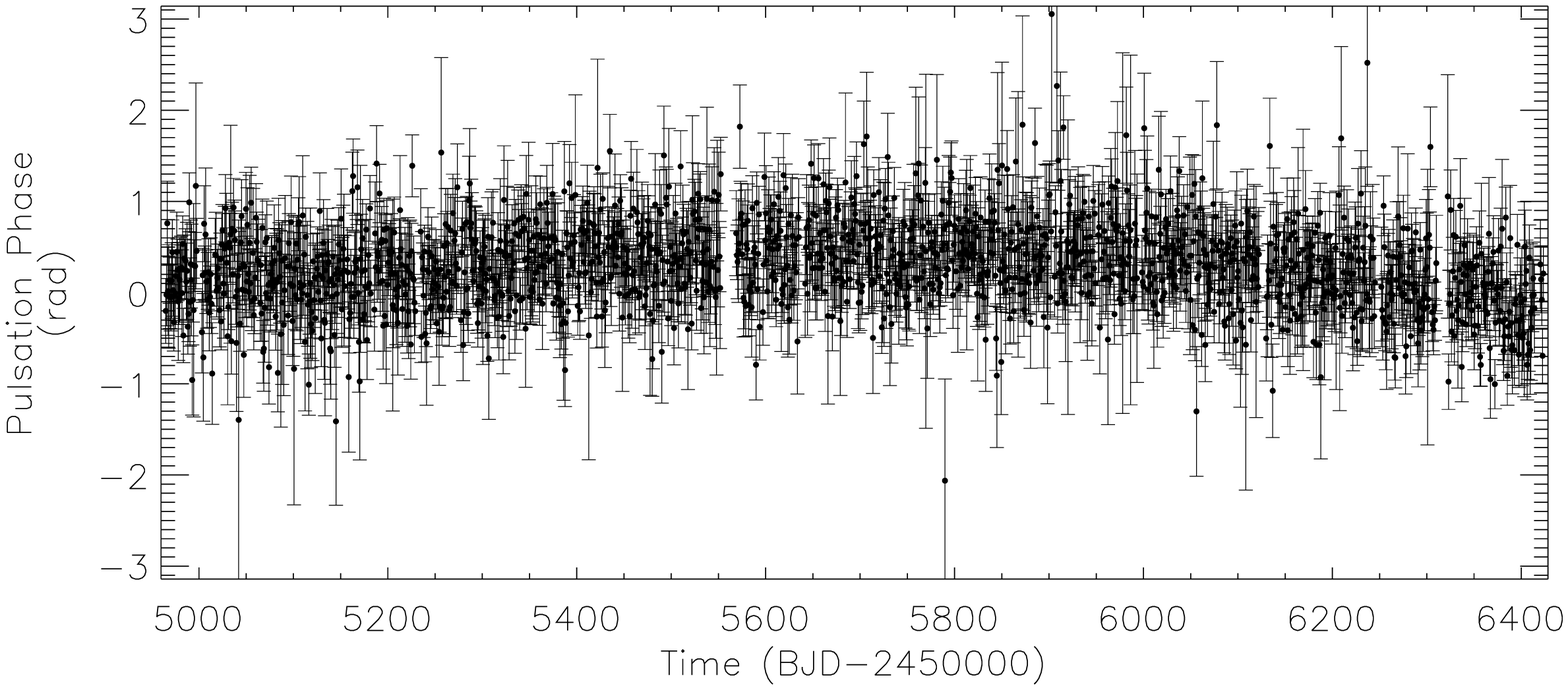}
    \includegraphics[width=0.5\textwidth]{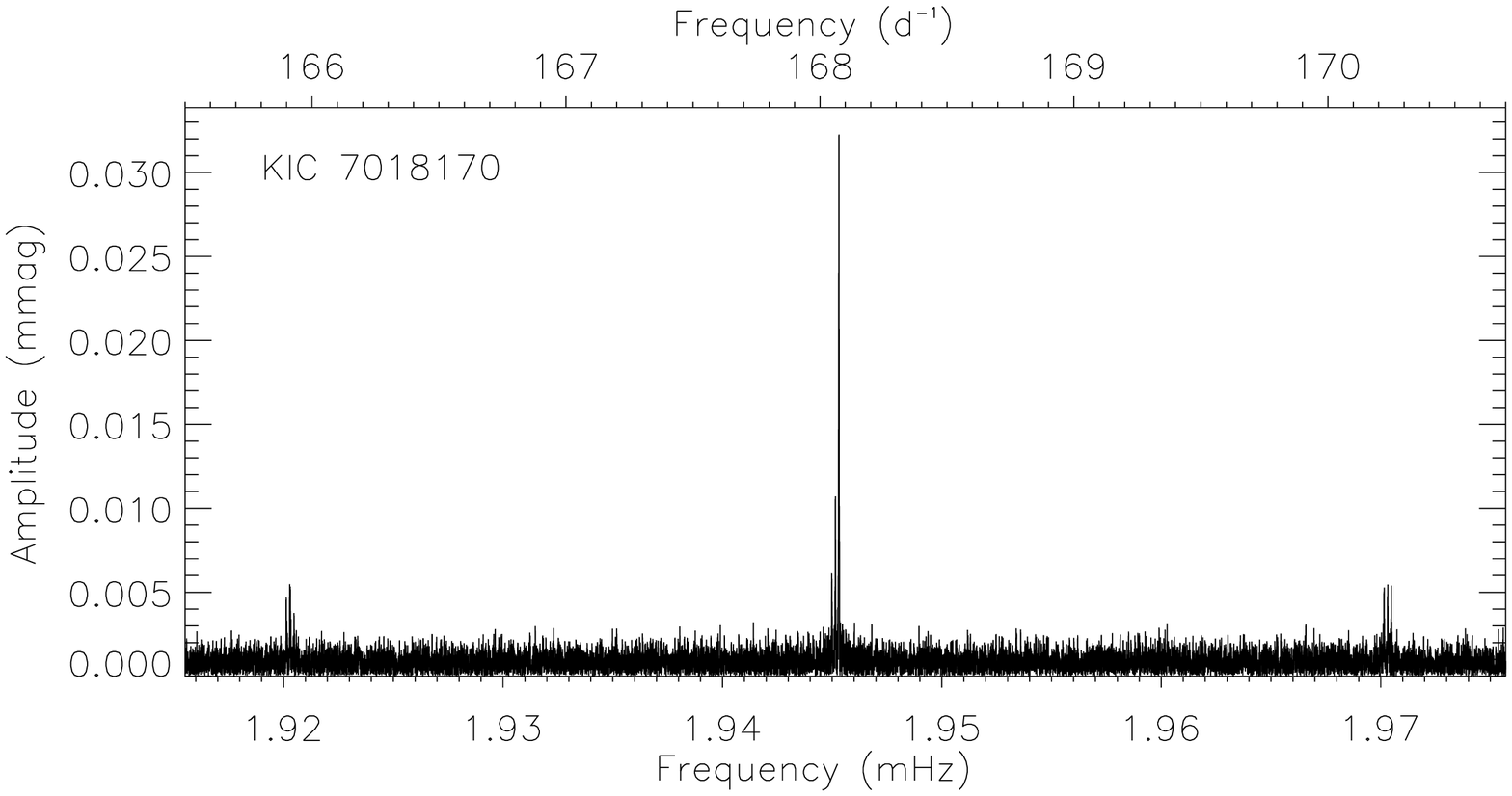}
    \includegraphics[width=0.5\textwidth]{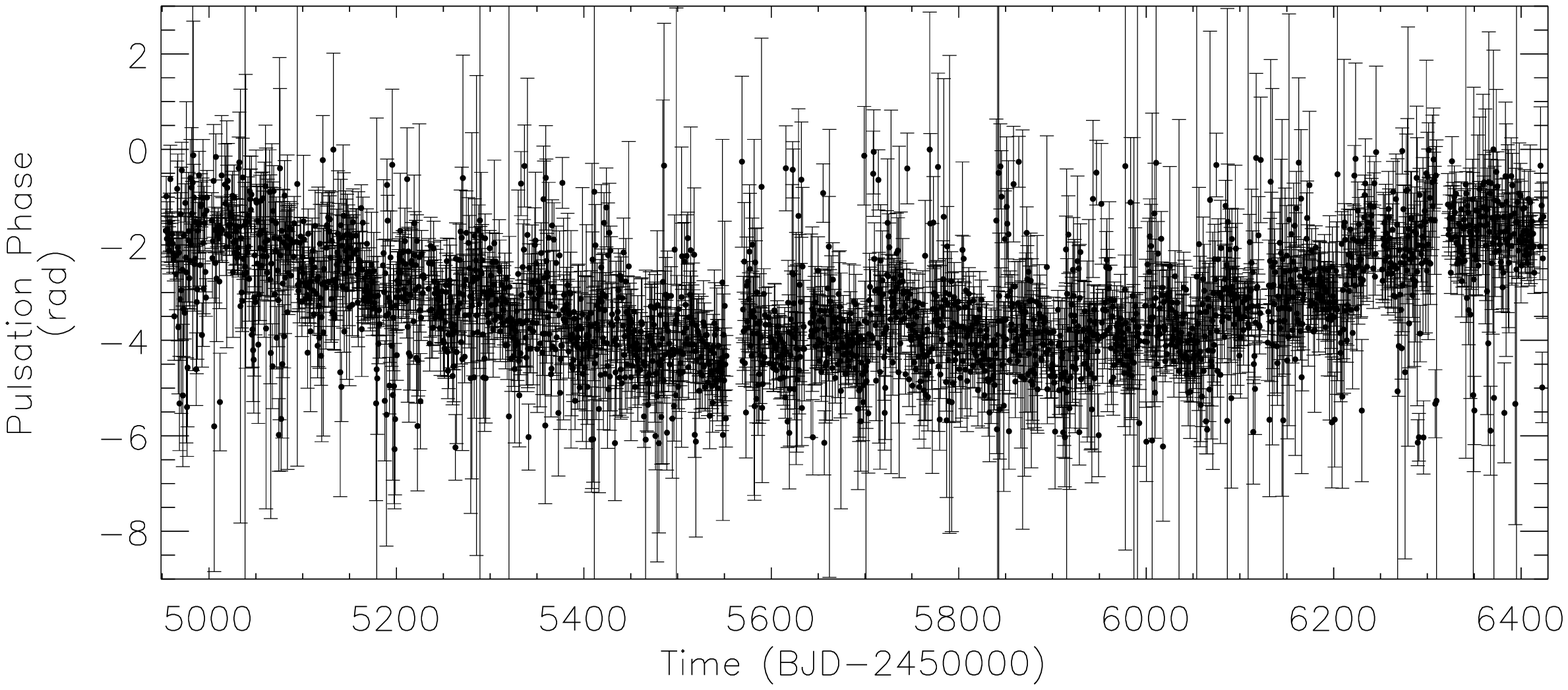}
    \includegraphics[width=0.5\textwidth]{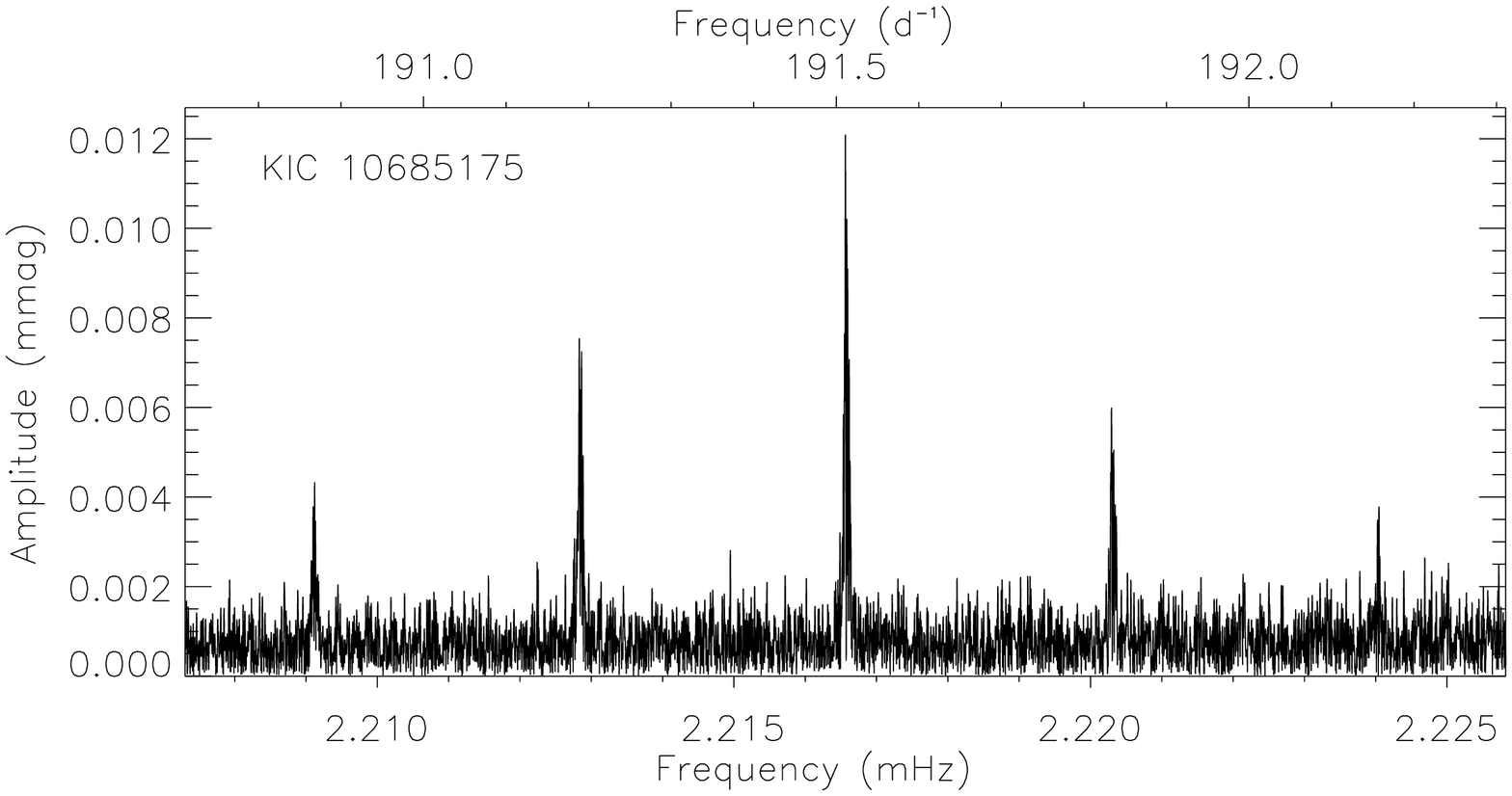}
    \includegraphics[width=0.5\textwidth]{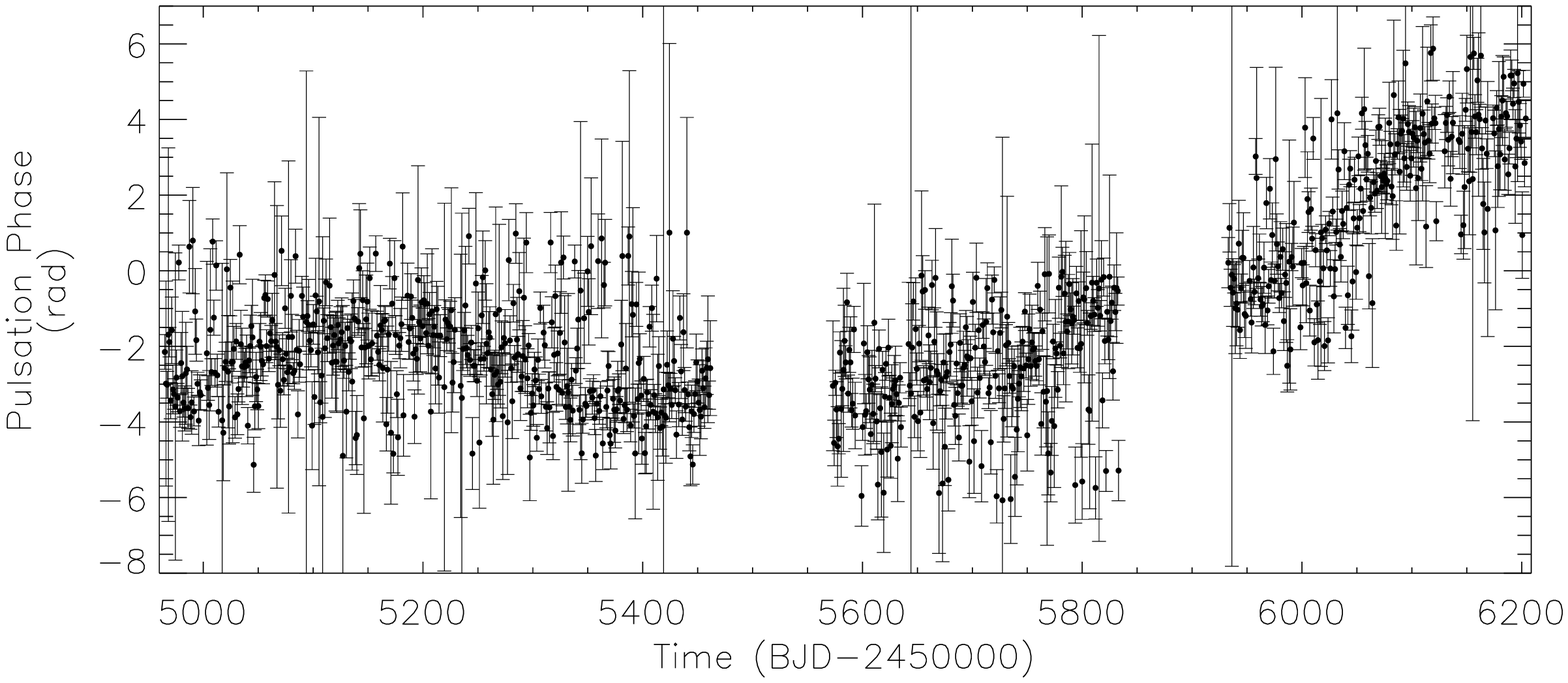}
    \includegraphics[width=0.5\textwidth]{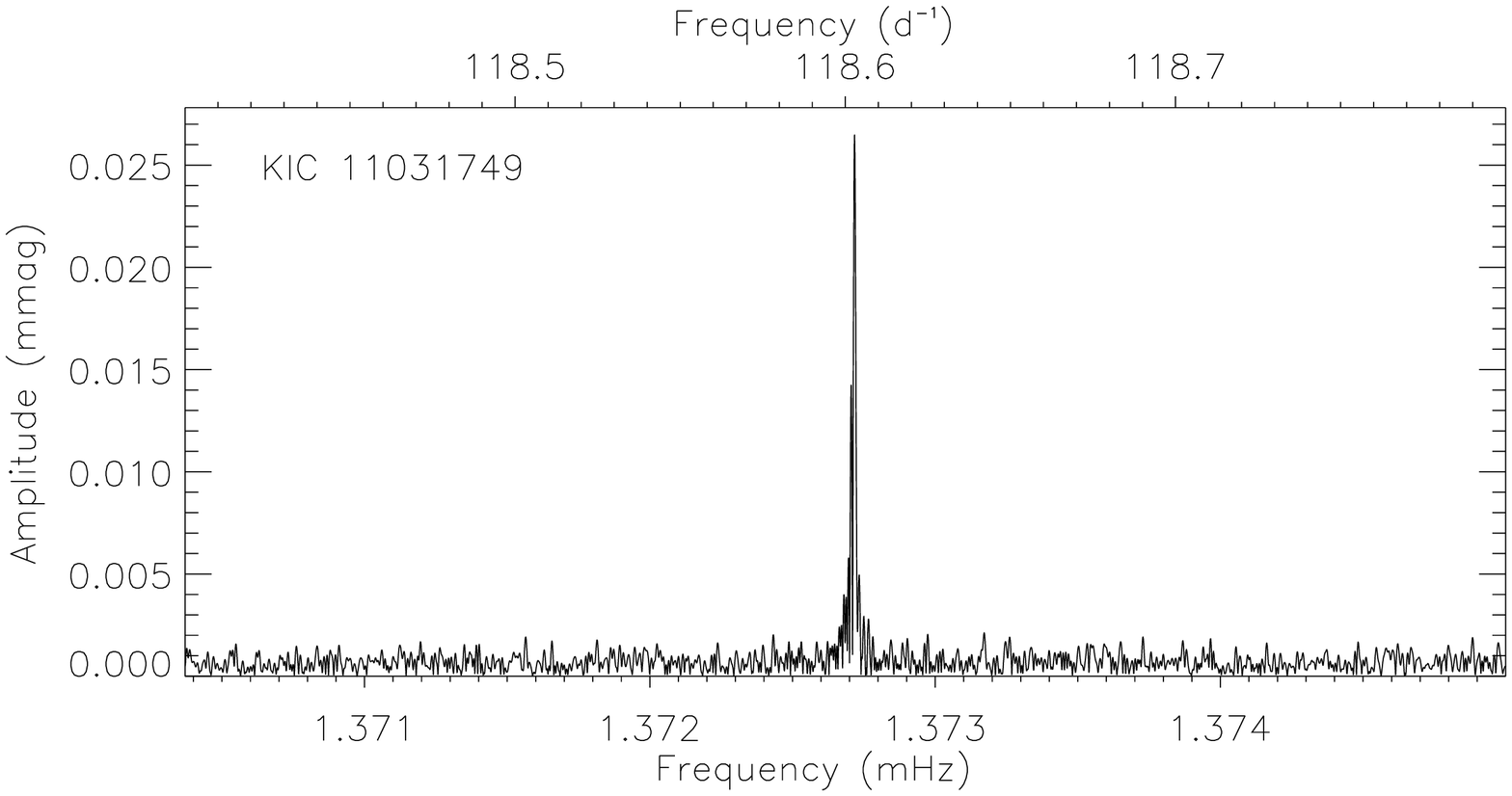}\hfill
    \includegraphics[width=0.5\textwidth]{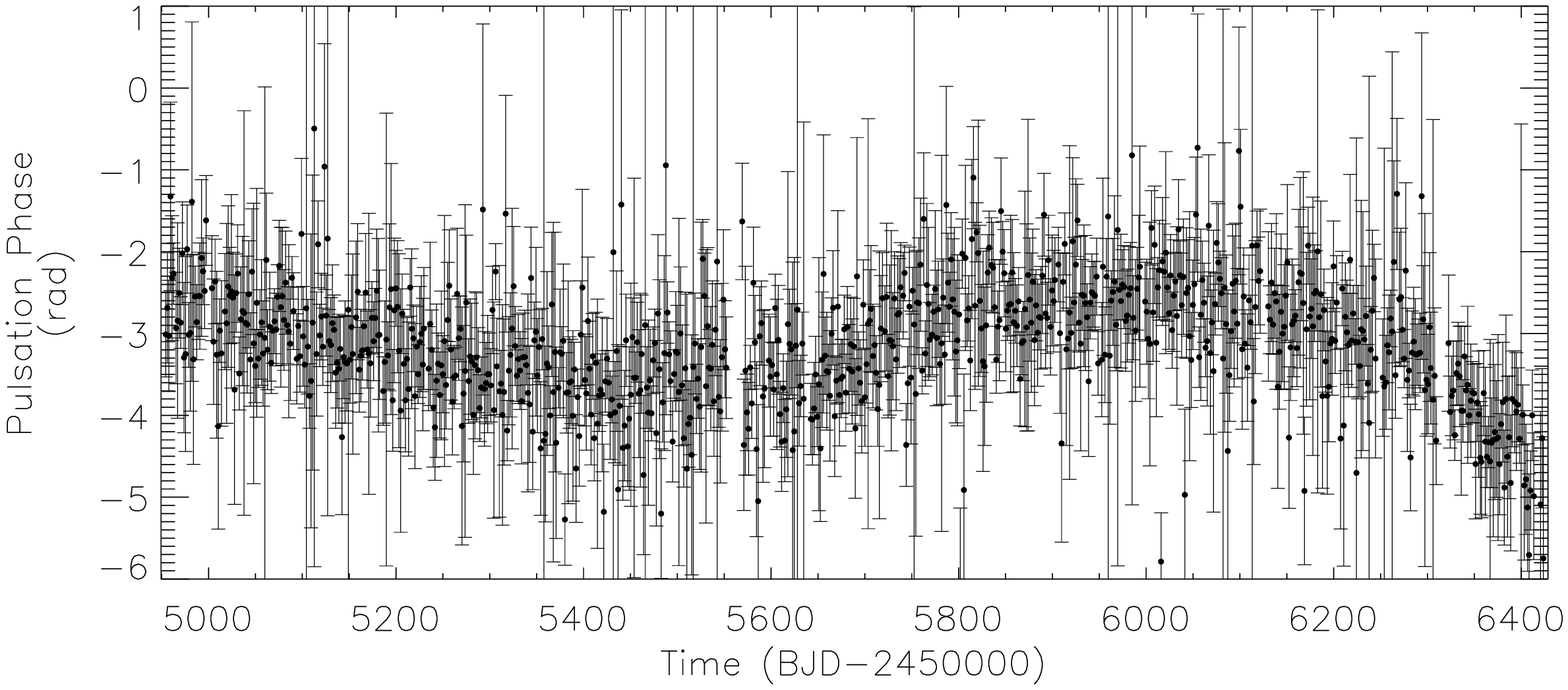}
    
    \caption{Each row in the figure corresponds to a different star. From top to bottom: KIC\,6631188, KIC\,7018170, KIC\,10685175 and KIC\,11031749. Left column: amplitude spectra of the LC light curve. Right column: phase variability of the principal mode in the star, as shown in the left column.}
    \label{fig:HEY_stars}
\end{figure}

\renewcommand{\thefigure}{\arabic{figure} (Cont.)}
\addtocounter{figure}{-1}

\begin{figure}
    \includegraphics[width=0.5\textwidth]{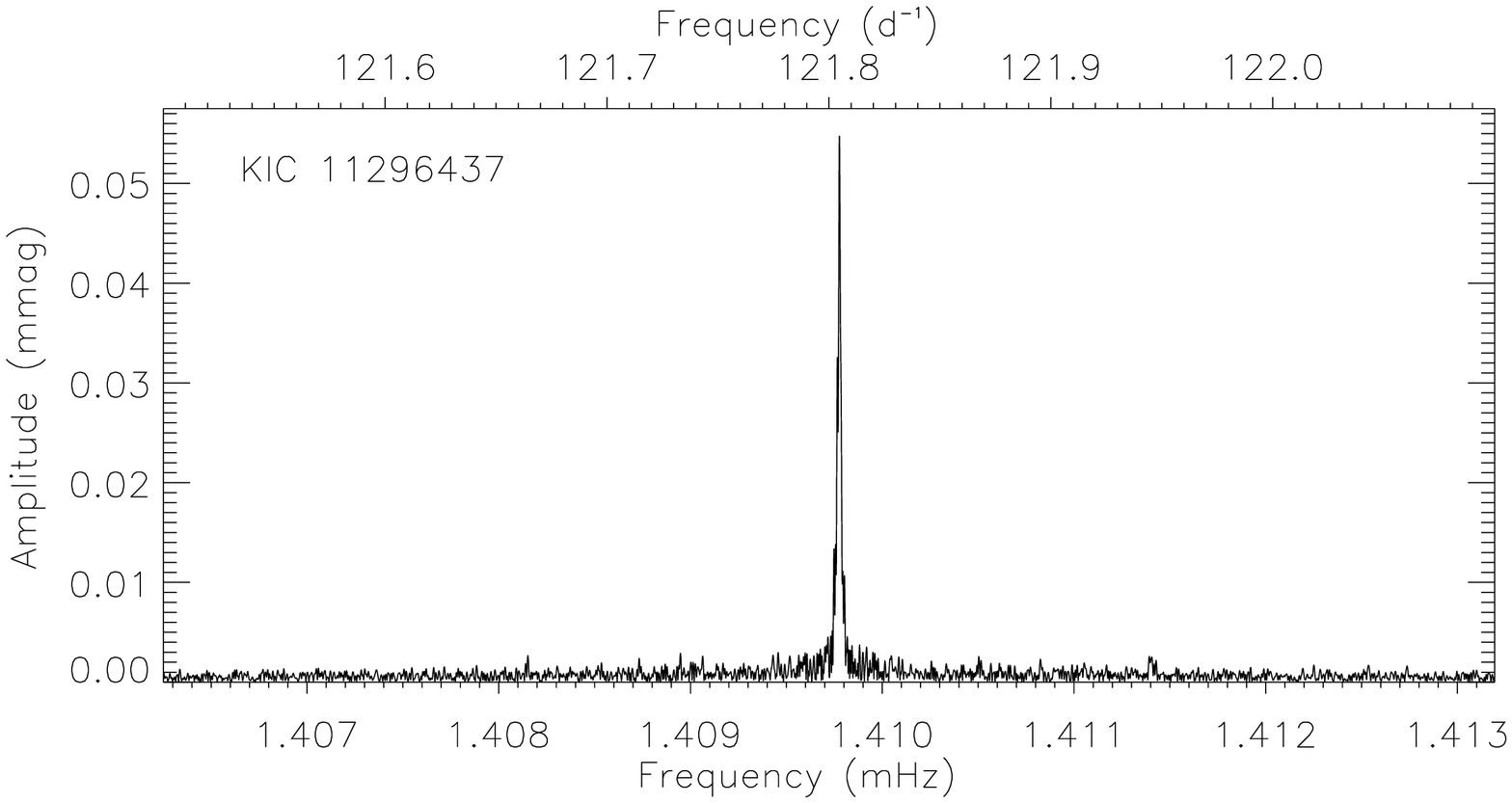}
    \includegraphics[width=0.5\textwidth]{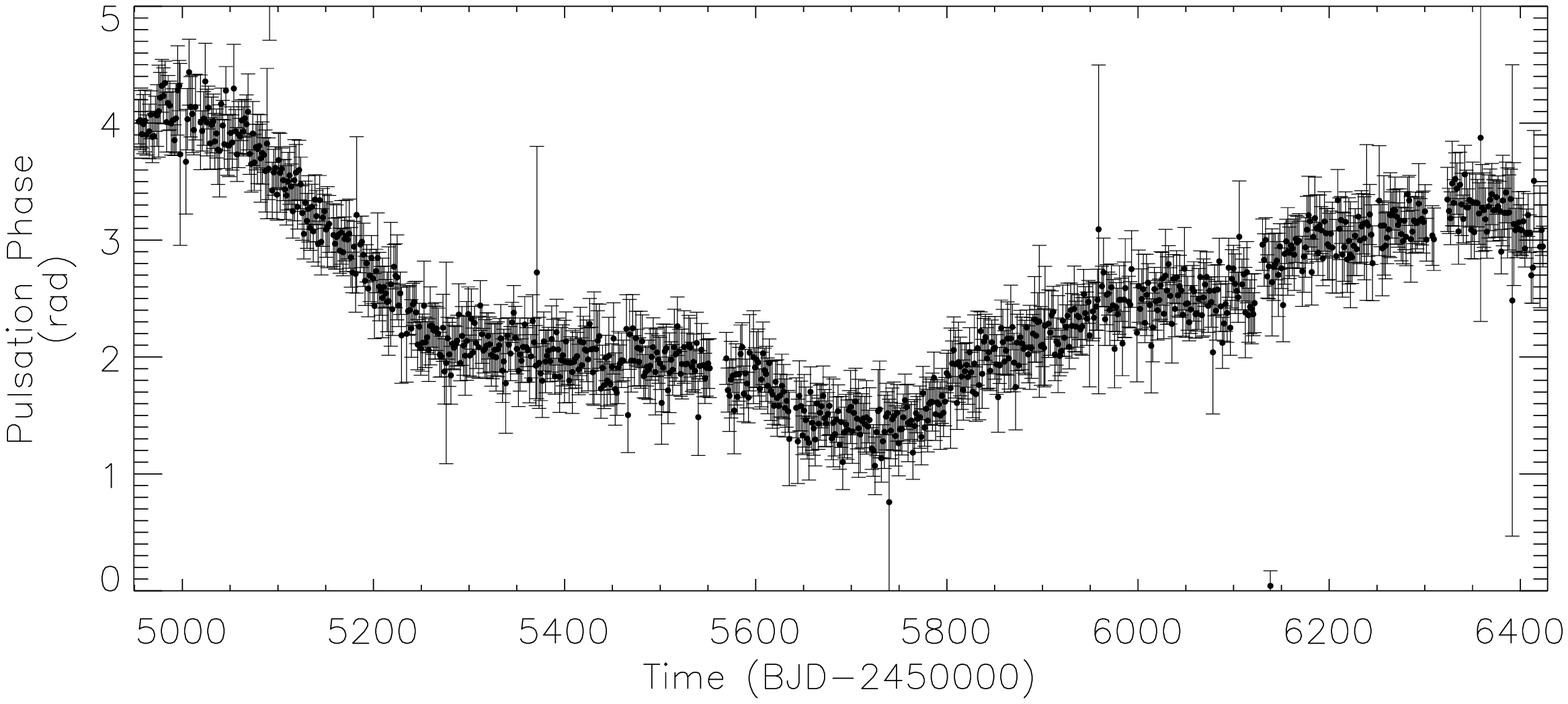}
    \includegraphics[width=0.5\textwidth]{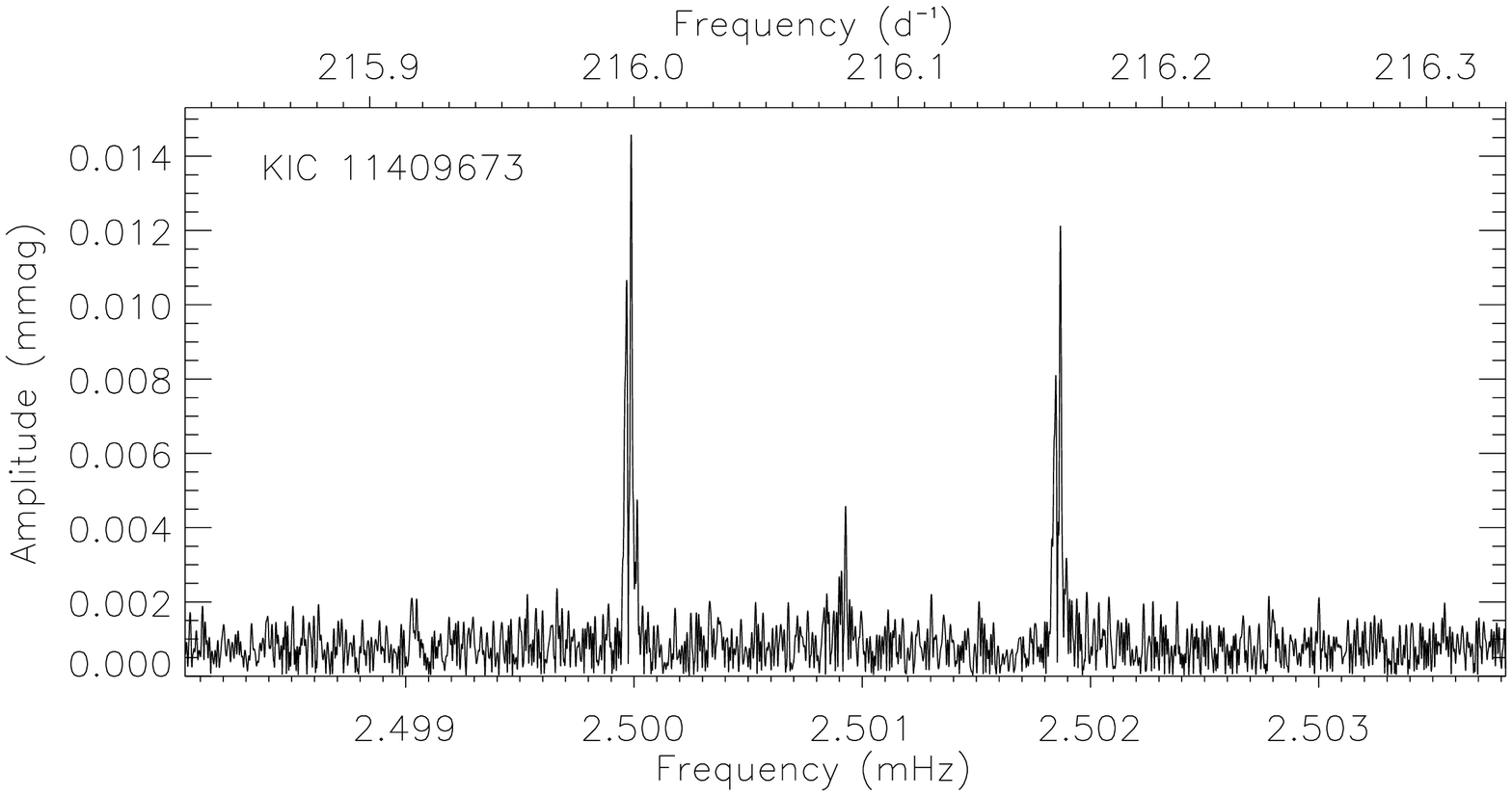}\hfill
    \includegraphics[width=0.5\textwidth]{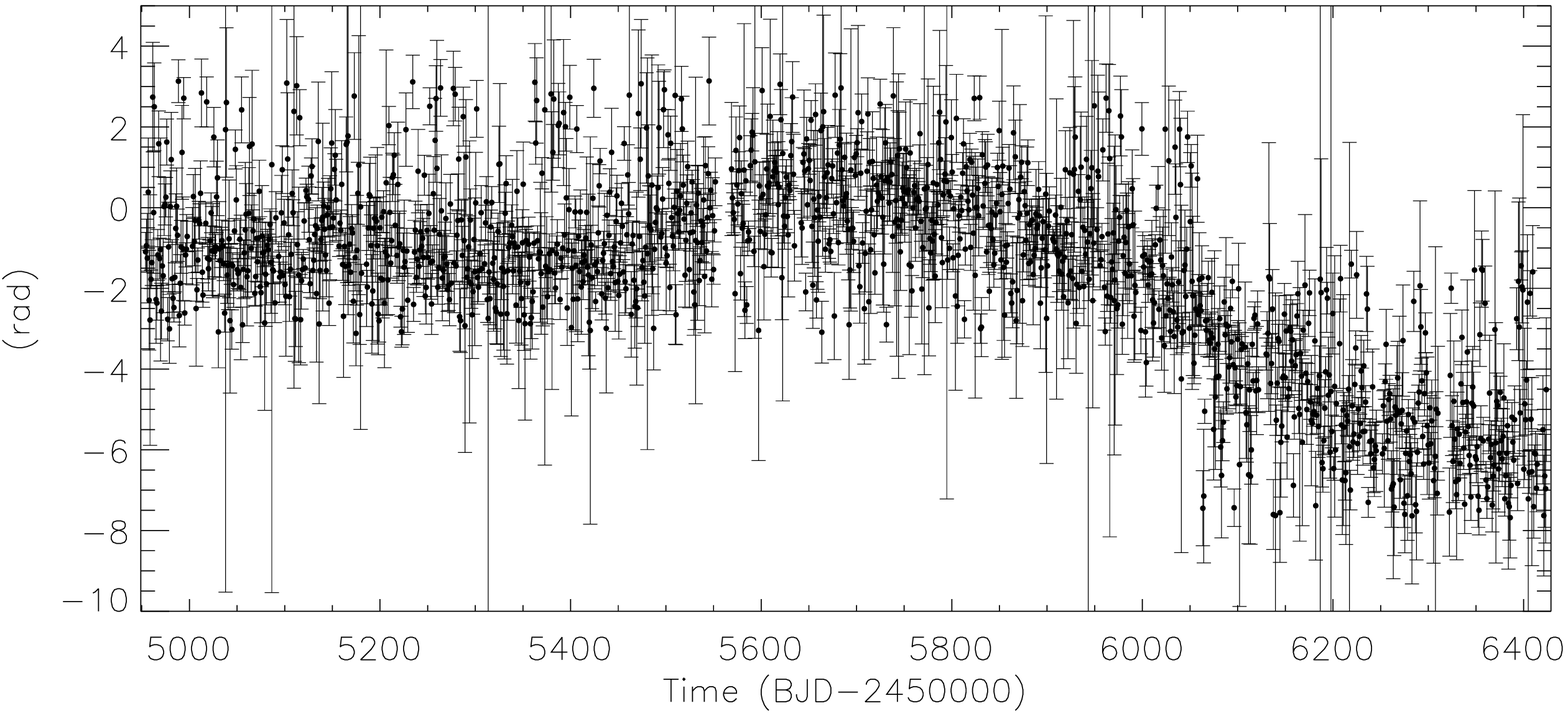}
    \caption{Continuation of Fig.\,\ref{fig:HEY_stars}. The top panels show KIC\,11296437 and the bottom panel shows KIC\,11409673.}
\end{figure}

\renewcommand{\thefigure}{\arabic{figure}}

Although these stars pulsate above the Nyquist frequency, as does KIC\,7582608, \citet{2019MNRAS.488...18H} was able to apply the oblique pulsator model to four of the six stars and found significant frequency variability in three of them. However, it seems that all six stars show a degree of frequency variability, as evidenced in the right column of Fig.\,\ref{fig:HEY_stars}. Again, the source of the frequency variability is unclear in these stars, but for KIC\,11031749 the change seems to be cyclic, but on the order of the length of the \kplr\ data set, making conclusions uncertain. Given the stellar parameters derived by \citet{2019MNRAS.488...18H}, two of the roAp stars show pulsations above the theoretical cut-off limit, an upper frequency limit where pulsations are not driven in models of non-magnetic stars of the same fundamental parameters \citep[e.g.,][]{1985PASJ...37..245S,1998MNRAS.301...31G,2011MNRAS.414.2576S}, thus providing more examples to challenge the theoretical models of these stars.

The objects discussed in this section will benefit from observations obtained with the ongoing {\it TESS} mission. Although data sets will be short, 2-min cadence observations will remove any alias ambiguities, as has been done for KIC\,10685175 \citep{2020ApJ...901...15S}, and amplitude suppression, potentially allowing for the detection of further modes in these stars, and thus a full asterosismic analysis. Furthermore, new, well separated in time, observations have the potential to shed light on the causes of the observed frequency/amplitude modulations observed by \kplr.\\

\subsubsection{Short Cadence observations}

There were 4 confirmed roAp stars observed in SC by the primary \kplr\ mission: KIC\,8677585, KIC\,10483436, KIC\,10195926 and KIC\,4768731. The initial publications on three of these stars were compiled with only a short section of the now complete 4-yr \kplr\ data, with KIC\,4768731 being the exception.

\citet{2011MNRAS.410..517B} published the first results of an roAp star observed by \kplr, namely KIC\,8677585, and provided a follow up study with more data in \citet{2013MNRAS.432.2808B}. The authors showed this star to be variable in two distinct frequency ranges through the identification of modes at 3.141\,\cd\ and 6.423\,\cd\ and many modes in the range $125-145$\,\cd, with harmonic and combination frequencies of the high-frequency group present. 

\begin{figure}
    \centering
    \includegraphics[width=0.5\textwidth]{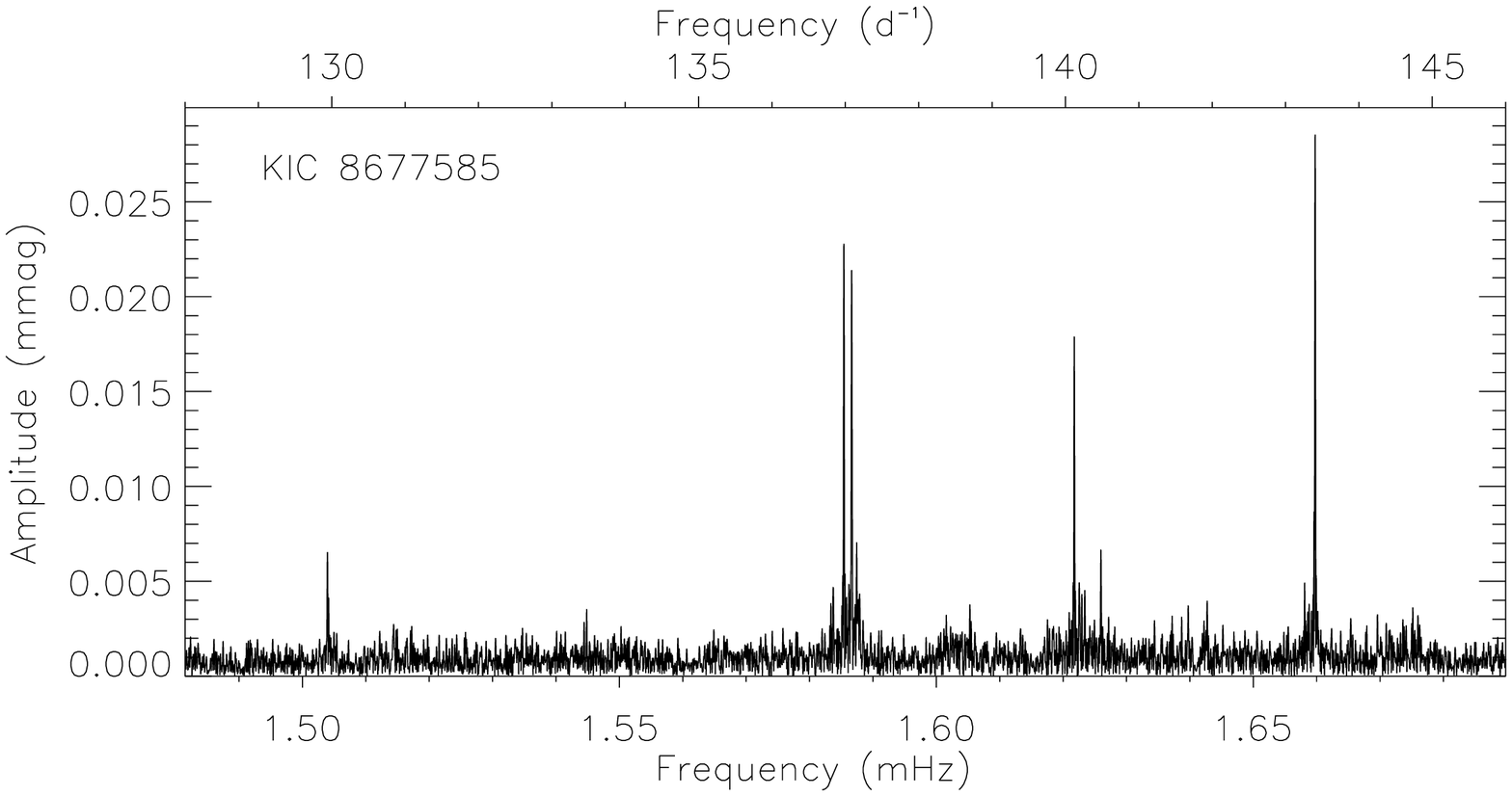}\hfill
    \includegraphics[width=0.5\textwidth]{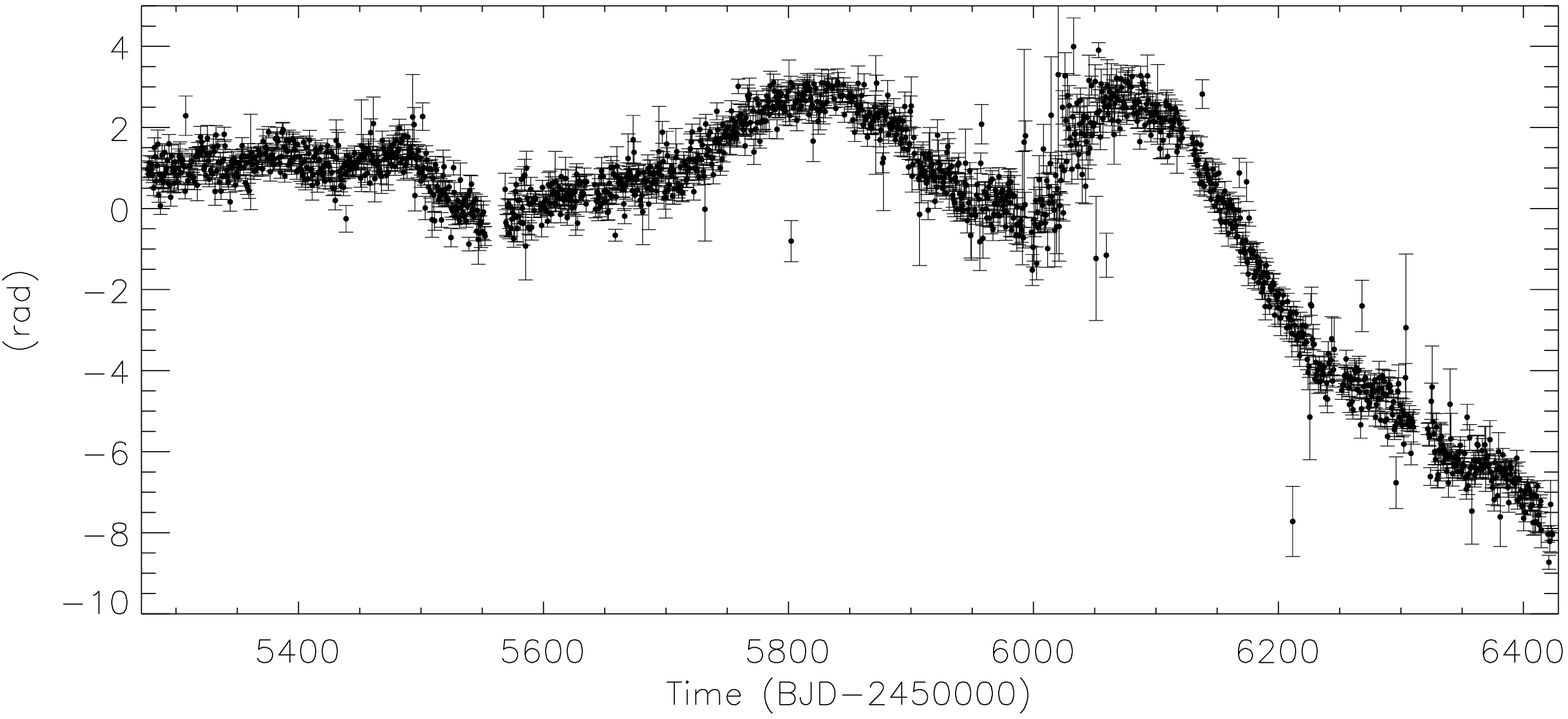}
    \caption{Left: the pulsations seen in KIC\,8677585 in the high-frequency range. Right: the phase variation of the principal mode in the left panels. There is significant frequency variability present in this mode.}
    \label{fig:K8677585}
\end{figure}

This star is among the group of roAp stars that show significant frequency and amplitude variability, as shown in Fig.\,\ref{fig:K8677585}, so precise mode identification becomes difficult. However, \citet{2013MNRAS.432.2808B} were able to measure the value of $\Delta\nu$ in this star to be $37.3\,\upmu$Hz which is close to the frequency of the long-period variation. Also linked to the low frequency modes, amplitude variability of some of the high-frequency modes occur with the same frequency, implying that these two phenomena are related. It was speculated that the low-frequency modes are g\,modes, but perhaps they are manifestations of the amplitude variations of the high-frequency modes, or a signature of beating.

The frequency variability for each of the observed modes in this star is also unique to each given mode. This is confirmation that the variability is intrinsic to the star, and not driven by an external body. This does then mean that all the pulsation cavities are changing over the observation period. Over the $\sim830$\,d of observations analysed by \citet{2013MNRAS.432.2808B}, none of the variability seems cyclic, as was suggested for HR\,3831 \citep{1994MNRAS.268..641K}. Are we then observing evolutionary changes in the star? Will revisiting this star in the future lead to a different $\Delta\nu$ determination? It is unclear at this point as to what can be inferred from these precise observations.

The second star to be published from the \kplr\ data was KIC\,10483436 by \citet{2011MNRAS.413.2651B}, with the analysis of a 27-d light curve. The number of harmonics of the rotation signature observed in this star was quite striking, indicating that with precise \kplr\ photometry, it is possible to observed small scale inhomogenities on the surface of stars, something that can later be investigated with high-resolution Doppler Imaging.  

\begin{figure}
    \centering
    \includegraphics[width=0.5\textwidth]{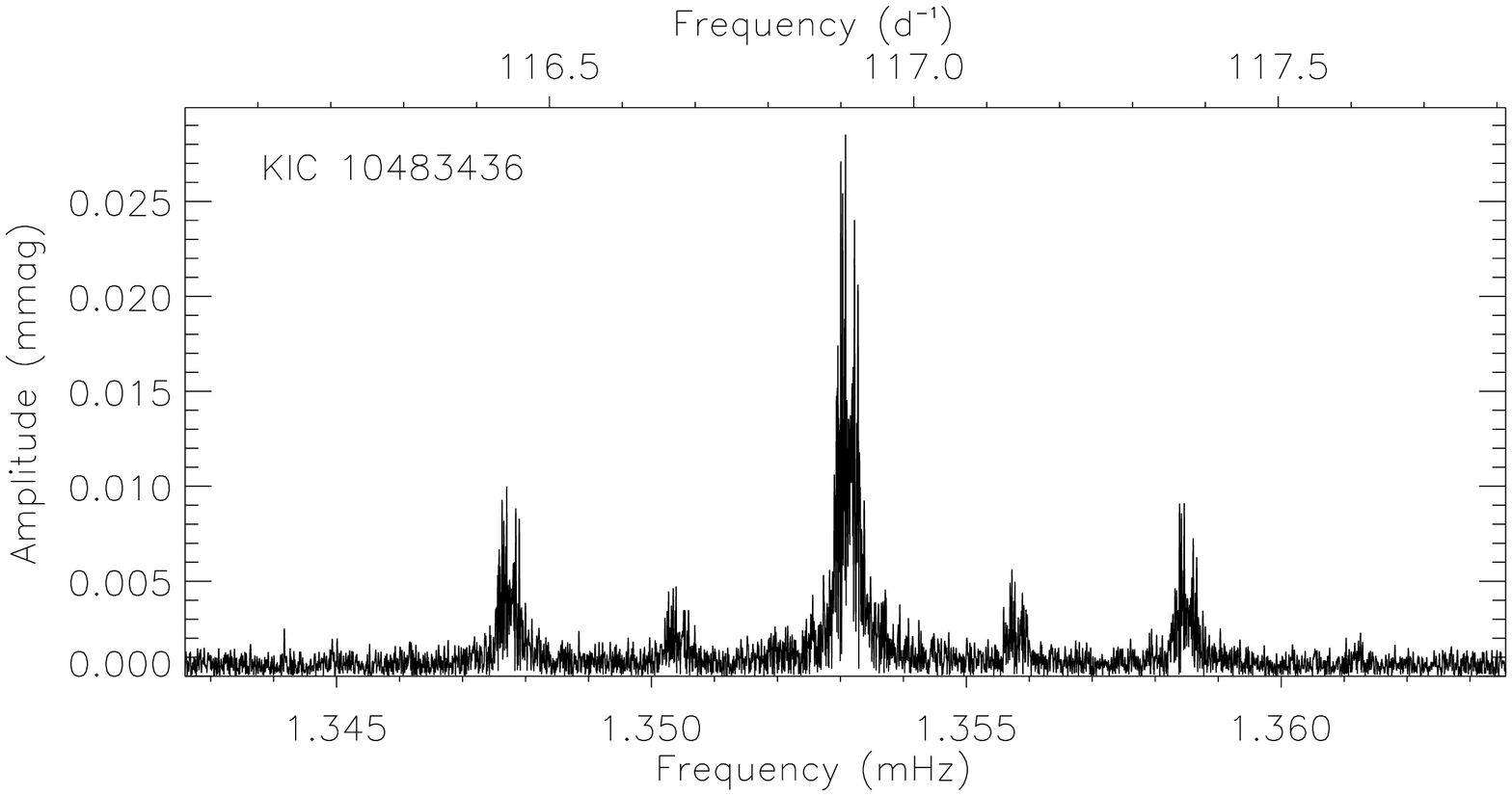}\hfill
    \includegraphics[width=0.5\textwidth]{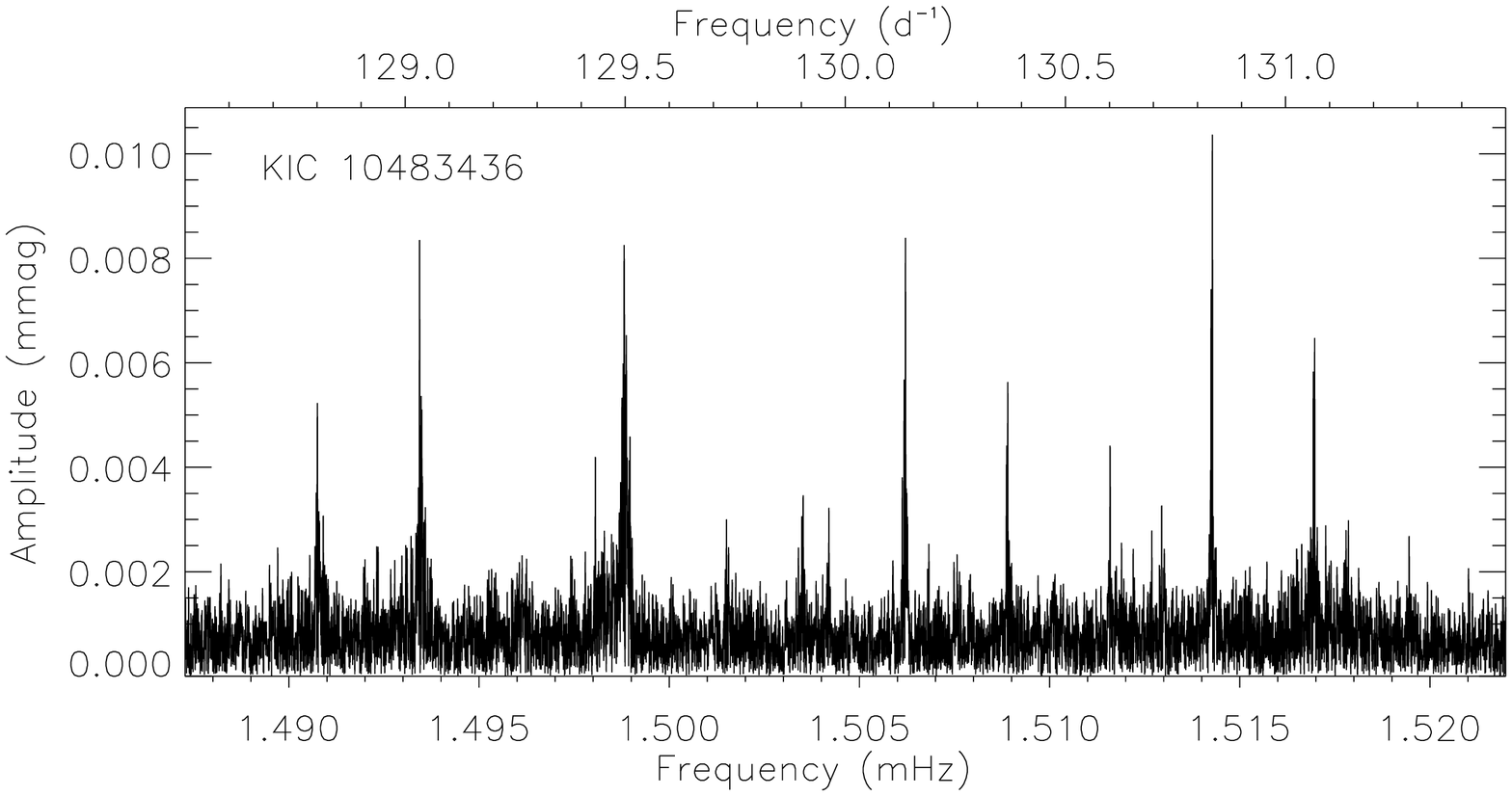}
    \includegraphics[width=1\textwidth]{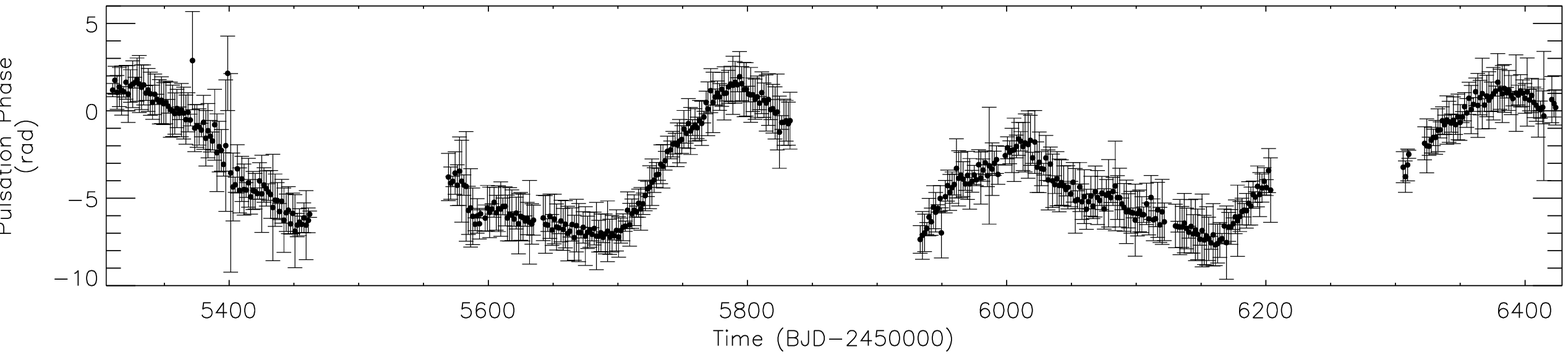}
    \caption{Top row: amplitude spectra of KIC\,10483436 showing the pulsation frequencies in this star. The left plot shows a quadrupole quintuplet with the peaks separated by the rotation frequency ($2.670\,\upmu$Hz,$P=4.303$\,d). The right panel shows many modes with rotational sidelobes, making mode identification less straightforward. Bottom: the phase variation of the principal mode in this star ($\sim1.353$\,$\upmu$Hz).}
    \label{fig:10483436}
\end{figure}

This star is clearly pulsating with a quadrupole mode with significant amplitude, and further modes at lower amplitude as shown in Fig.\,\ref{fig:10483436}. Although the discovery paper cites only 2 modes in this star, it is clear from an investigation using the full \kplr\ data set for this star, that many more modes are present, forming a clump around the low-amplitude mode previously reported. The identification of the number of modes in this star is hampered by the rotation sidelobes caused by oblique pulsation, causing modes and sidelobes to overlap in frequency. An independent determination of the rotation frequency, maybe with ground-based multicolour data since the amplitude is dependent on colour \citep[e.g.,][]{1996MNRAS.280....1K,2017MNRAS.471.3193D}, could allow this to be untangled.

Although not reported in the discovery paper, with the additional data available, it is evident that there is significant frequency variability in the principal pulsation mode in this star, with indications also in the low-amplitude modes. This variability is non-cyclic and shows sudden changes, implying that the changes are caused by internal phenomena rather than an external source such a companion which would introduce regular, smooth changes in the pulsation phase \citep[see examples in][]{2018MNRAS.474.4322M}. This is therefore another example that the precise, long-term, monitoring by \kplr\ has revealed information that would otherwise have gone undetected.

KIC\,10195926 was reported as an roAp star by \citet{2011MNRAS.414.2550K}, with the analysis of 25\,d of SC data. In the low-frequency regime, they identified a sub-harmonic of the rotation frequency which has an unknown origin. That feature is still present in the longer data set now available, but the source is still not explained. Spectroscopic observations of the star would be needed to determine if this frequency is related to a binary companion. However, a more likely suggestion by \citet{2011MNRAS.414.2550K} is that the sub-harmonic signature is that of an r\,mode -- a global Rossby wave that is driven by the radial component of vorticity interacting with the Coriolis force \citep[for a detailed discussion of r\,modes see][]{2018MNRAS.474.2774S}. This is the first roAp star, and indeed Ap star, that is thought to host an r-mode oscillation. The visibility of r\,modes is dependent on inclination, spot size and contrast ratio and stellar rotation rate, with slow rotation posing significant visibility issues. The availability of high-precision, long time-base \kplr\ observations has the power to enable the detection of these signatures. Now, with more data, a full investigation into this possibility is possible. 

\begin{figure}
    \centering
    \includegraphics[width=0.5\textwidth]{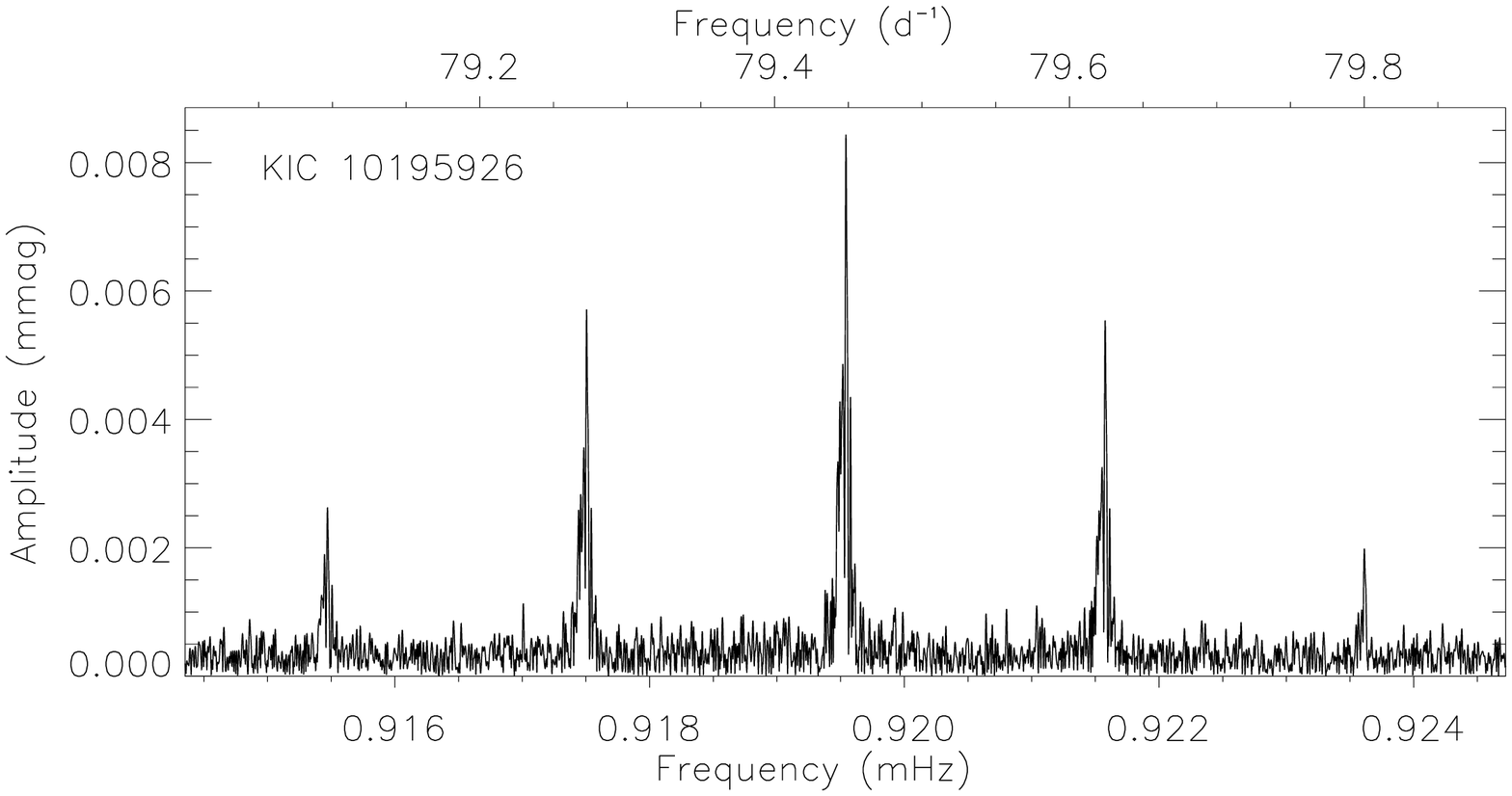}\hfill
    \includegraphics[width=0.5\textwidth]{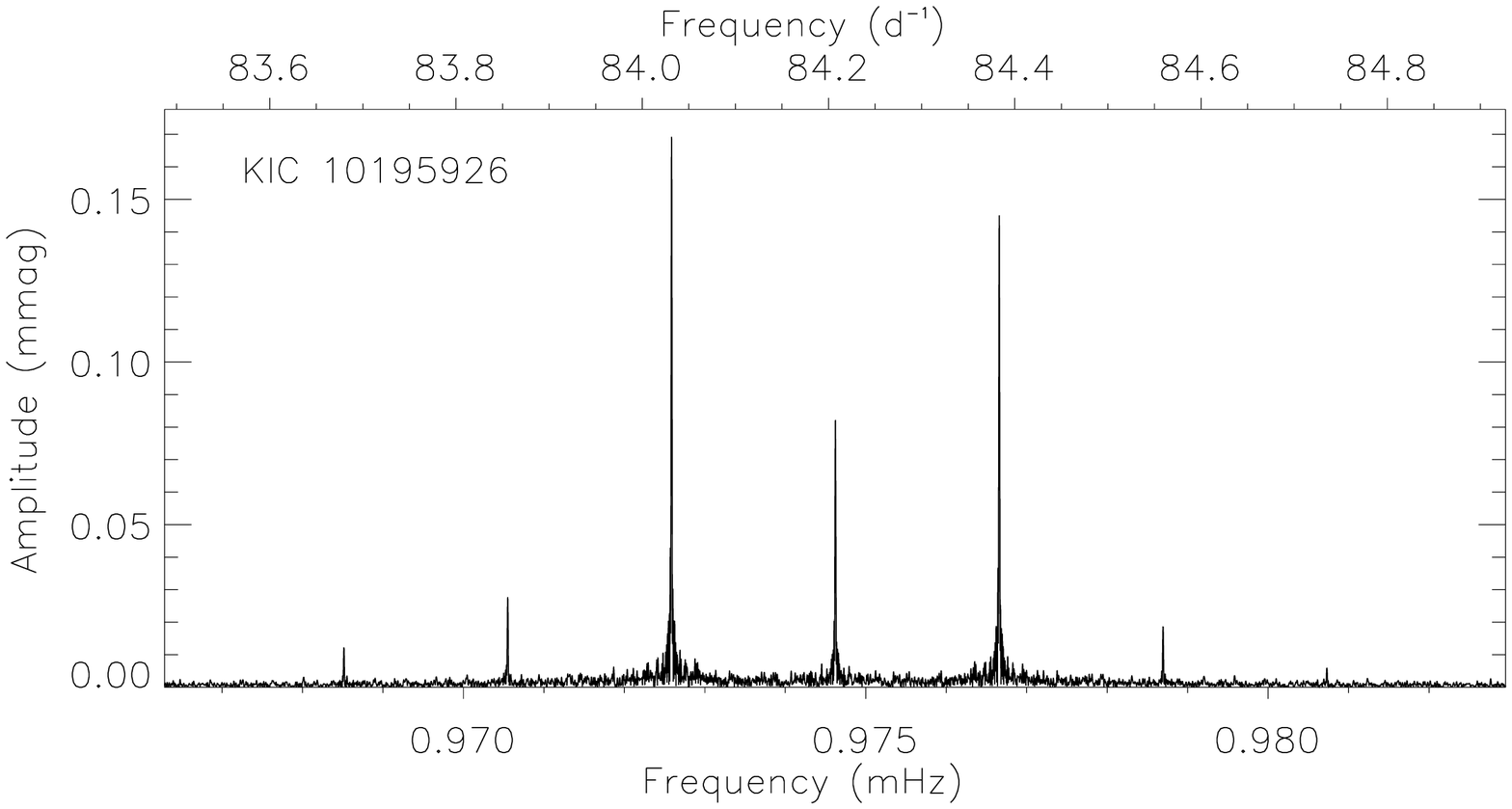}
    \includegraphics[width=0.5\textwidth]{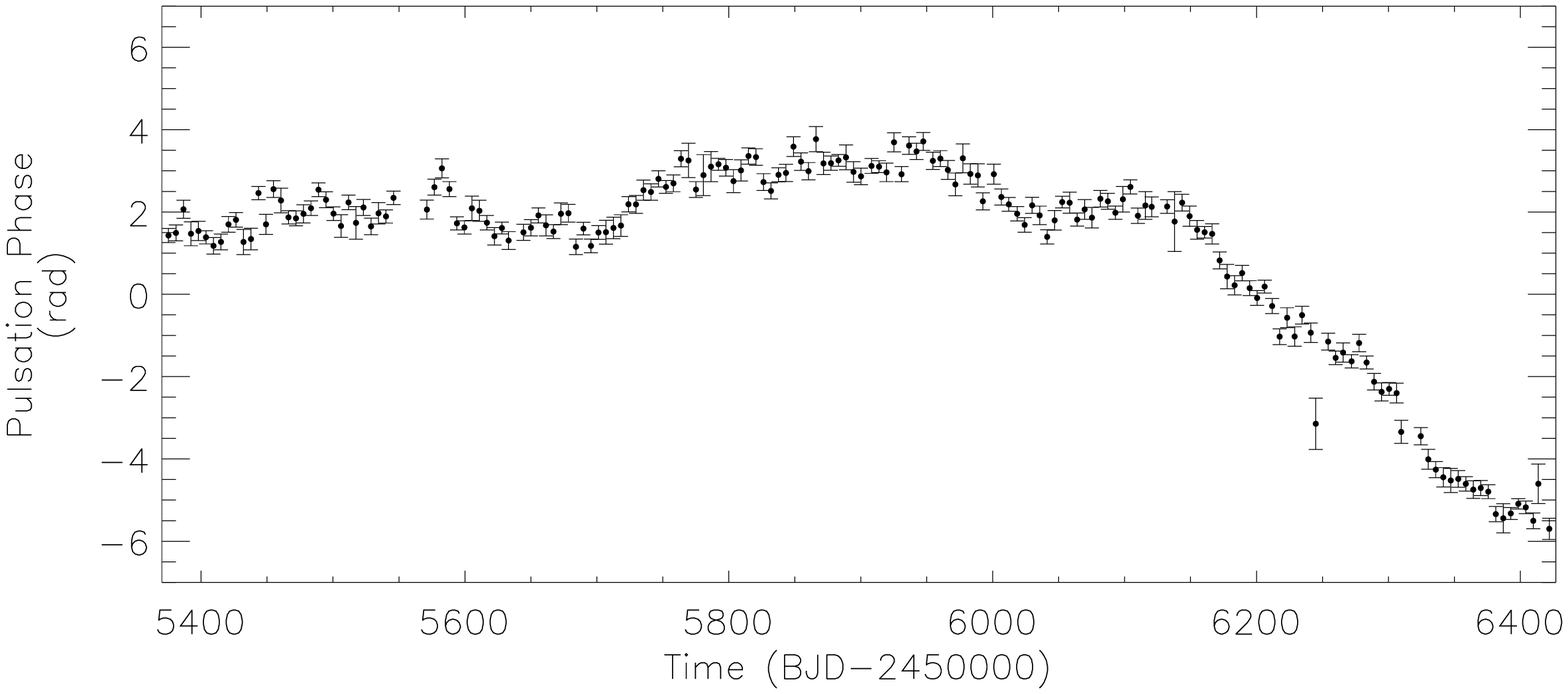}\hfill
    \includegraphics[width=0.5\textwidth]{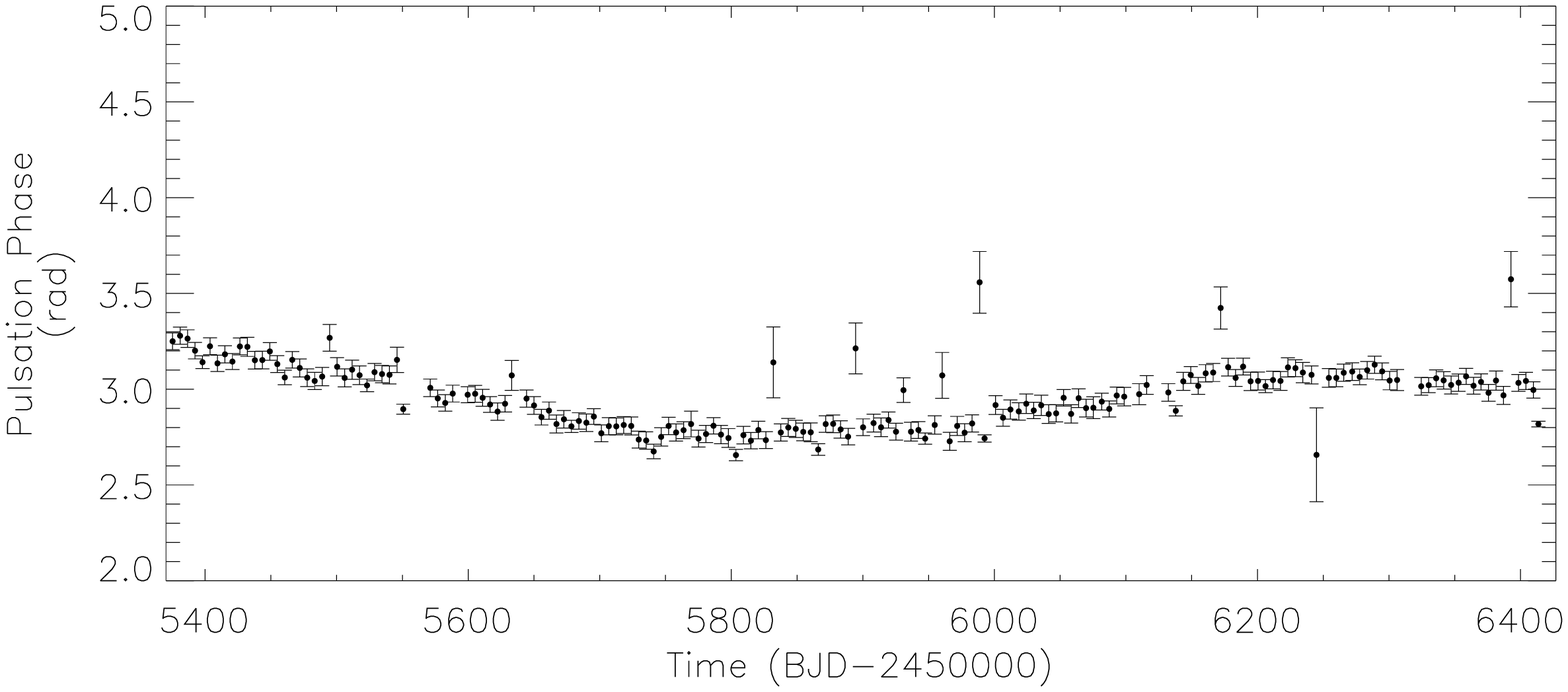}
    \caption{Top row: amplitude spectra of the two pulsation modes in KIC\,10195926. The left plot shows that the lowest frequency mode is split into a quintuplet with the extra data from later \kplr\ quarters. In both cases, the sidelobes are split by the rotation frequency ($2.036\,\upmu$Hz, $P=5.685$\,d). Bottom row: corresponding phase variations of the pulsation modes in the top row. The high-frequency mode on the right has a fairly stable frequency over 4 years, but the low-frequency mode has a significant phase change.}
    \label{fig:10195926}
\end{figure}

There were two pulsation modes identified in KIC\,10195926, one mode split into a septuplet and one into a triplet by oblique pulsation and distortion (Fig.\,\ref{fig:10195926}). From the analysis of the phase variations of the pulsations over the rotation cycle, it was concluded that both modes are $\ell=1$ dipole modes; the triplet arises from a pure mode dipole mode, while the septuplet represents a distorted mode. However, since the relative sidelobe amplitudes for each mode differed, it was proposed that there are two pulsation axes in this star, and thus the geometry of the modes is different. The cause of this is proposed to be the interaction of the magnetic field and rotation on the pulsations. It has been shown by \citet{2002A&A...391..235B} that the difference in obliquity angle between two consecutive modes should be small in most cases, with \citet{2000MNRAS.319.1020C} and \citet{2004MNRAS.350..485S} showing that at specific frequencies, the magnetic field can greatly affect the modes. Now, with more data available, the low frequency mode is actually split into a quintuplet, thus the problem of the multiple axes in this star should be revisited. 

Furthermore, in light of new results, we propose another interpretation of these different relative amplitudes. Recent observations by the {\it TESS} mission of HD\,6532 show a distinctly different multiplet shape than the $B$-photometric observations presented by \citet{1996MNRAS.280....1K} (see \citet{2020ASSP...57..313K} for comparison plots). New ground-based multicolour observations confirm this difference in mode multiplet structure over 5 filters (Holdsworth et al., in prep.). Since each filter probes a different atmospheric depth, a simple comparison of data from different filters cannot be made, and strong conclusions about the mode geometry cannot be drawn. With KIC\,10195926, the wide \kplr\ passband that probes a wide range of atmospheric layers, coupled with the significantly stratified atmospheres in Ap stars and potentially different mode sensitivities at a certain atmospheric depth, may result in the different mode structures seen in the two modes in KIC\,10195926, rather than the presence of two distinct pulsation axes. This, however, is conjecture with a detailed theoretical study required to definitively solve this conundrum.

A preliminary look at this much longer data set reveals KIC\,10195926 to be yet another frequency variable star. Interestingly in this case, the principal mode is only slightly variable, and perhaps in a cyclic way, but the low-amplitude mode has a very different, and more significant, variation in its frequency. With this different variability for each mode, it further complicates the physical interpretation of this star.

\begin{figure}
    \centering
    \includegraphics[width=0.49\textwidth]{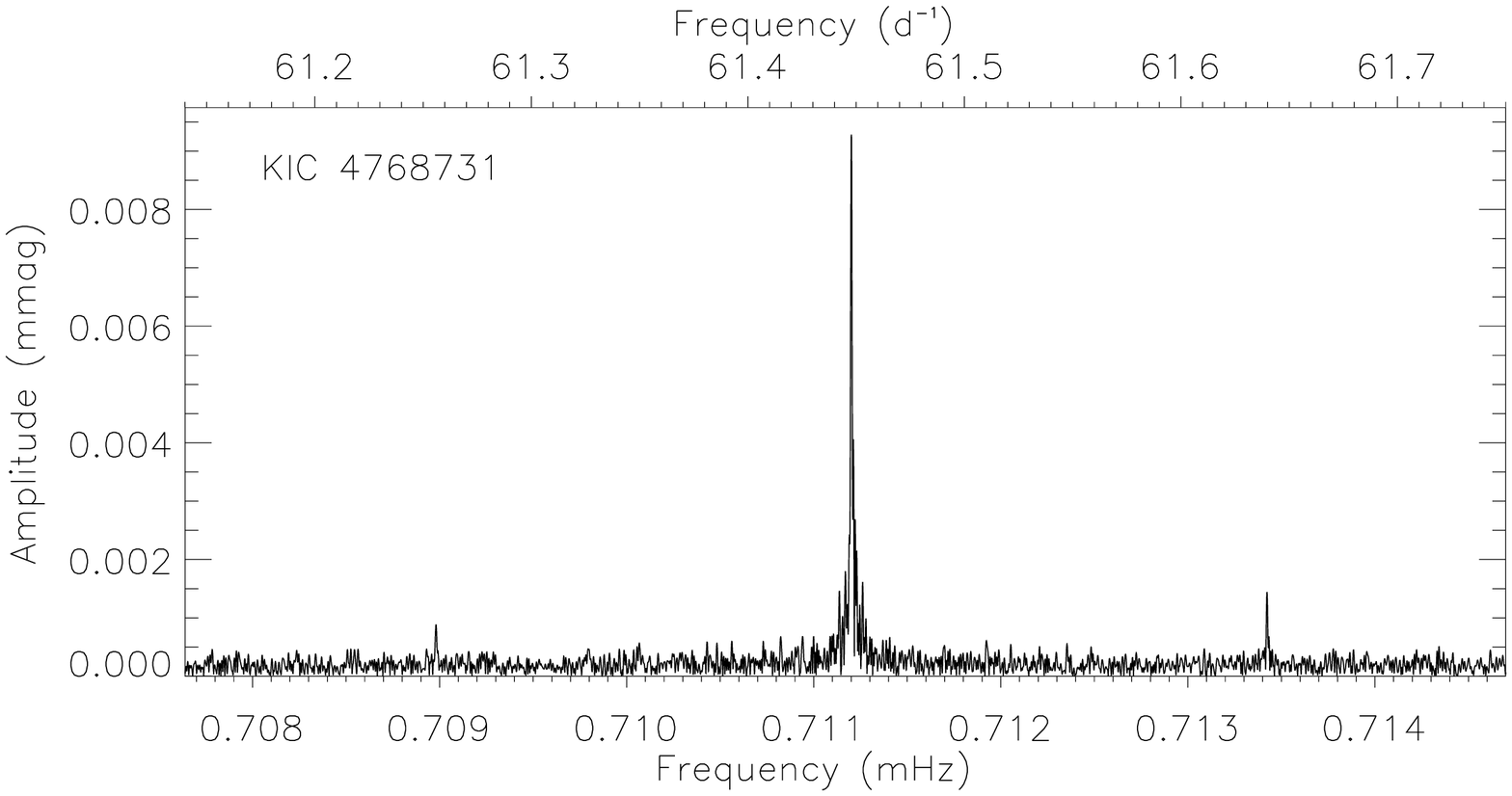}
    \includegraphics[width=0.49\textwidth]{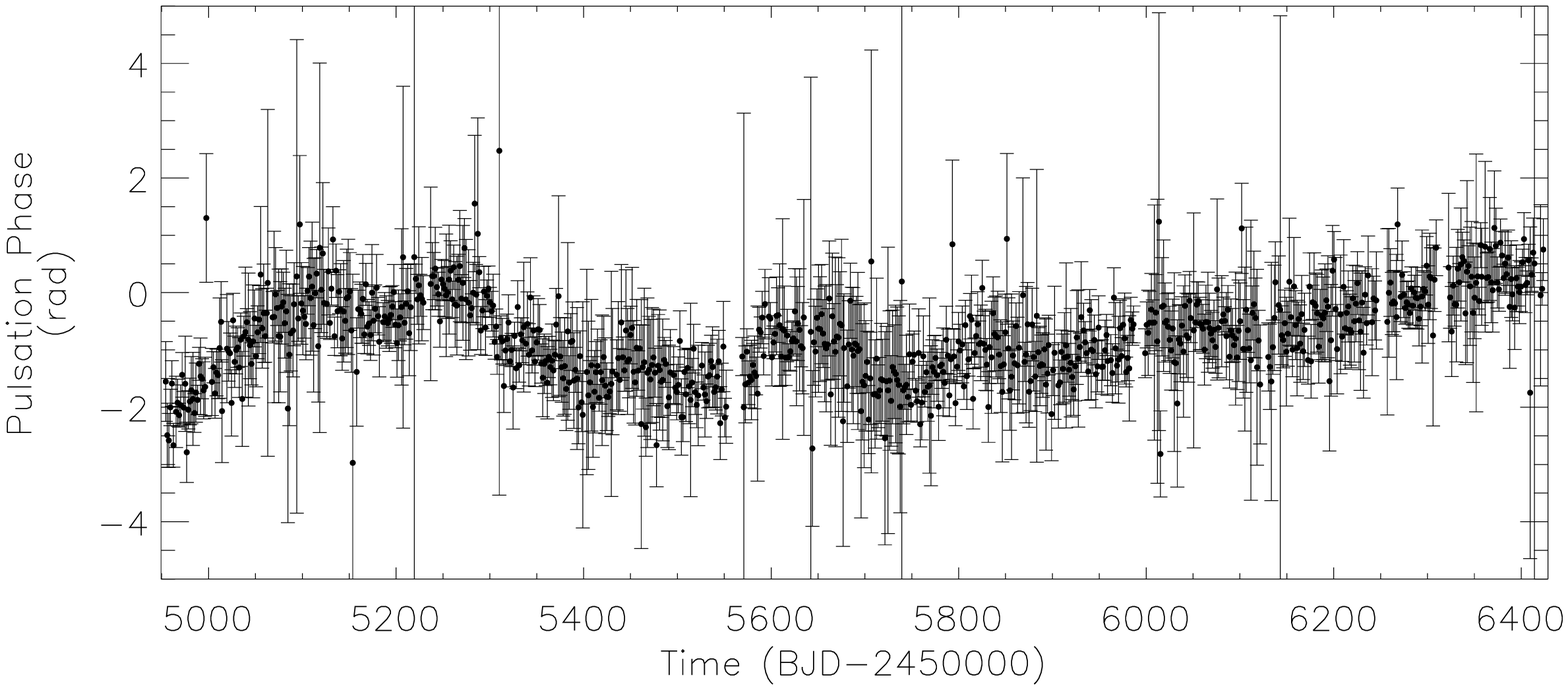}\hfill
    \caption{Left: amplitude spectrum of the LC data for KIC\,4768731. The pulsation mode and two rotationally split, by $2.2218\,\upmu$Hz ($5.209$\,d), sidelobes are evident. Right: phase variation of the pulsation mode, showing frequency variation in this star.}
    \label{fig:4768731}
\end{figure}

The final star to be observed in SC mode in the primary \kplr\ mission was KIC\,4768731. This star was discovered to be an Ap star by \citet{2015MNRAS.450.2764N}, with \citet{2015MNRAS.452.3334S} providing an analysis of the pulsation behaviour. Unfortunately, KIC\,4768731 was only observed for a single month in SC mode, but does have a full set of LC observations. The SC data allowed for the discovery of a rotationally split triplet in this star, which is understood to be a dipole mode under the oblique pulsator model. As with many of the roAp stars discussed here, KIC\,4768731 shows signs of frequency variability in the LC data set (Fig.\,\ref{fig:4768731}).

Given the lack of SC data for this star, not much further information could be gained from its light curve. Spectroscopically though, this star shows only weak over abundances of rare earth elements. When considering the numbers of Ap stars in clusters, \citet{2009AJ....138...28A} was able to estimate the age when peculiarities in Ap stars become strong; for a star the mass of KIC\,4768731, this age is about 12\,\% of its total main sequence life time. Therefore, KIC\,4768731 may provide an opportunity to revisit the links between age, chemical peculiarities and pulsation in the roAp stars. \\

\subsection{K2 Mission}

With the failing of a second reaction wheel needed for precise pointing, the \kplr\ Space Telescope was reconfigured into the K2 mission \citep{2014PASP..126..398H}. With only two functioning reaction wheels, the spacecraft was balanced against the solar radiation pressure with thruster firings. This configuration provided the opportunity to observe new parts of the sky in high-precision and short cadence and was a welcome change for the observations of roAp stars. Despite the 6\,hr occurrence of the thruster firings required to maintain precision pointing, these K2 data allowed the detailed analysis of three previously known roAp stars. Despite searches of the K2 data \citep[e.g.,][]{2018A&A...616A..77B}, no new roAp stars have yet been confirmed in the K2 data.\\

\subsubsection{Long Cadence observations}

Only one roAp star was observed in LC mode in the K2 mission: HD\,177765. This star was identified as an roAp star through the analysis of time-resolved UVES spectroscopic observations \citep{2012MNRAS.421L..82A}, after a null detection in ground based photometric $B$ observations \citep{1994MNRAS.271..129M}. At the time of discovery, this star showed the longest period pulsation mode, at 23.6\,min.  

HD\,177765 was observed in campaign 7 in LC mode. The data were analysed by \citet{2016IBVS.6185....1H}, where it was found that the star is a multi-periodic roAp star, with three independent modes. %(see Fig.\,\ref{fig:177765})
The separation of the modes is not consistent with theoretical predictions of the large frequency separation, so an in-depth study could not be presented for this star. The largest separation, at $\sim1.3$\,\cd, also explains the reporting by \citet{2012MNRAS.421L..82A} of only a single mode, given their short data set.

%\begin{figure}
%    \centering
%    \includegraphics[width=0.5\textwidth]{HD_177765_ft.eps}
%    \caption{Amplitude spectrum of the LC data for HD\,177765. Three modes are clearly present.}
%    \label{fig:177765}
%\end{figure}

In the white-light \kplr\ data, the highest amplitude mode has an amplitude of $11.0\pm0.8$\,$\upmu$mag. This is much below the ground-based detection limit, even when considering the amplitude when converting to the $B$-band. It is understandable that \citet{1994MNRAS.271..129M} did not detect the variability in this star. Given the low amplitude and the short data set for this star, it is not possible to draw any conclusions in the search for frequency variability in this star.\\

\subsubsection{Short Cadence observations}

Two roAp stars were observed in SC mode by K2: HD\,24355 and 33\,Lib (HD\,137949). HD\,24355 had only ground-based photometric survey data available prior to the K2 observations \citep{2014MNRAS.439.2078H}, while 33\,Lib had been extensively studied with both ground-based photometry and spectroscopy \citep[e.g.,][and references therein]{1991MNRAS.249..468K,2011MNRAS.416.2669S}.

HD\,24355 was observed in campaign 4, with the data analysed by \citet{2016MNRAS.462..876H}. They found just a single pulsation mode in the star, but with {\it 13} rotationally split sidelobes. The presence of so many sidelobes is a result of significant distortion of the pulsation mode. With the presence of four high-amplitude sidelobes, the authors concluded that the star was pulsating in a quadrupole mode, and were able to model the amplitude variation to that effect.

%\begin{figure}
%    \centering
%    \includegraphics[width=1\textwidth]{PW-K2_data.eps}
%    \caption{The pre-whitening steps in the search for the rotational sidelobes in HD\,24355. They are split by the rotation frequency of the star ($0.4146\,\upmu$Hz, $P=27.9158$\,d). Note the changing amplitude scale from row-to-row.}
%    \label{fig:24355}
%\end{figure}

The pulsation frequency in HD\,24355 is much higher that the theoretical upper limit, given the spectroscopic constraints. The observed frequency is $224.304$\,\cd\ whereas the cut-off frequency is about 164\,\cd. This brings into question the driving mechanism for this star. It is unclear whether the significant distortion of the mode is related to its super-critical nature. Without the K2 observations, the distortion would probably not have been detected, given that the amplitudes of the extended sidelobes do not exceed 40\,$\upmu$mag in the \kplr\ passband.

33\,Lib was observed during campaign 15 of the K2 mission. These data, analysed by \citet{2018MNRAS.480.2976H}, revealed a much more complex pulsation signature than was previously seen in either photometric or spectroscopic observations. The K2 data confirmed the presence of three modes as detected from the ground, but allowed for the detection of 11 independent modes. However, beyond these 11 modes, there are still signatures of variability in the light curve, as evidenced by excess power in an amplitude spectrum of the residuals. The source of the excess power was not investigated by the authors, but given that the pulsation mode frequencies in 33\,Lib are close to the theoretical cut-off frequency, there may be some excitation of short-lived modes by turbulent pressure. However, this is conjecture, and requires investigation.

\begin{figure}[htb]
    \begin{minipage}{0.34\linewidth}
    \caption{Top: schematic view of the pulsations in 33\,Lib. Note the logarithmic amplitude scale. Bottom: schematic view of the pulsations around the harmonic. Note that the separation between the modes is the same in the two plots, and not twice in the bottom plot as one would expect.}
    \label{fig:33Lib}
    \end{minipage}
    \hspace{0.06\linewidth}
    \begin{minipage}{0.55\linewidth}
        \includegraphics[width=\linewidth]{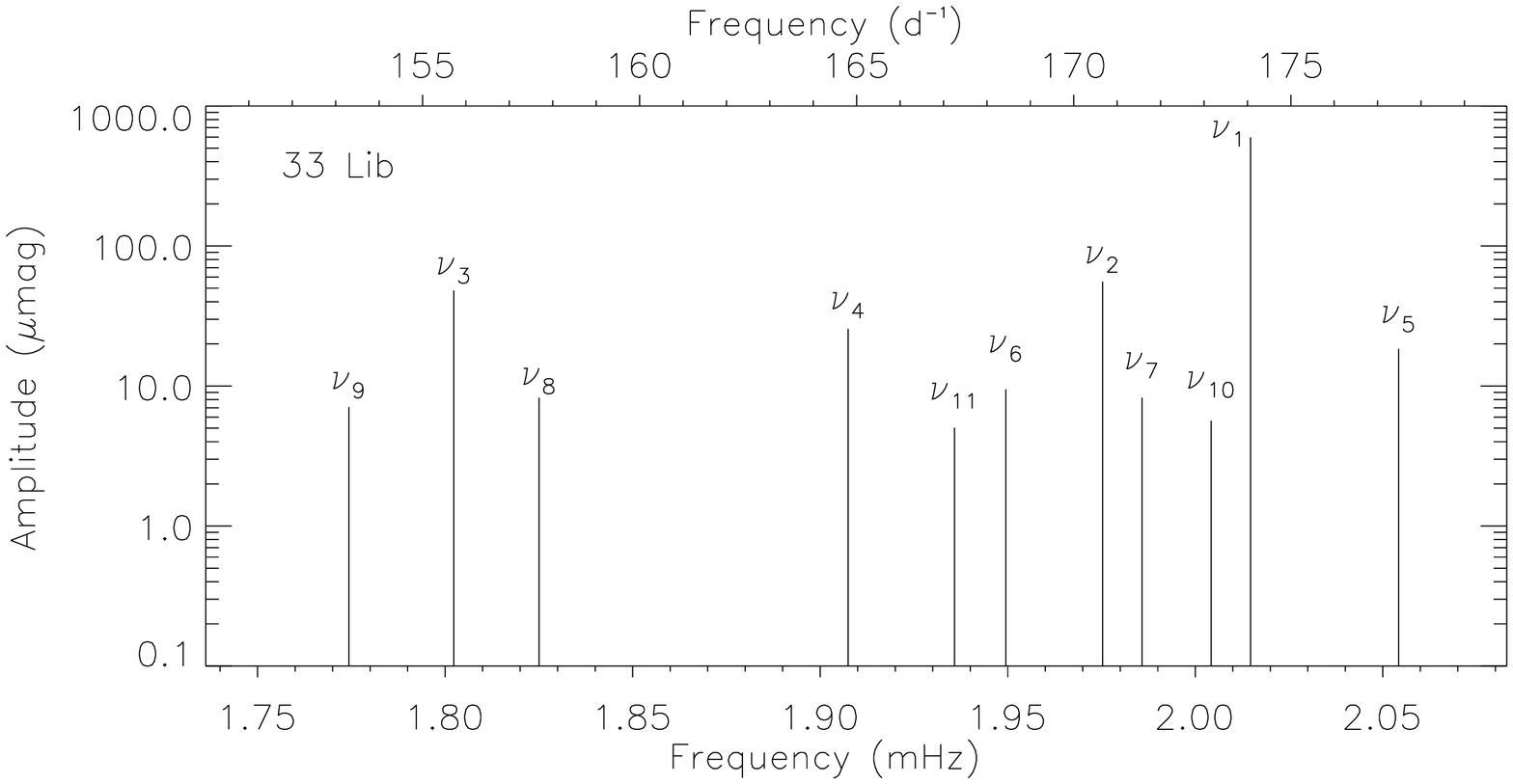}
        \includegraphics[width=\linewidth]{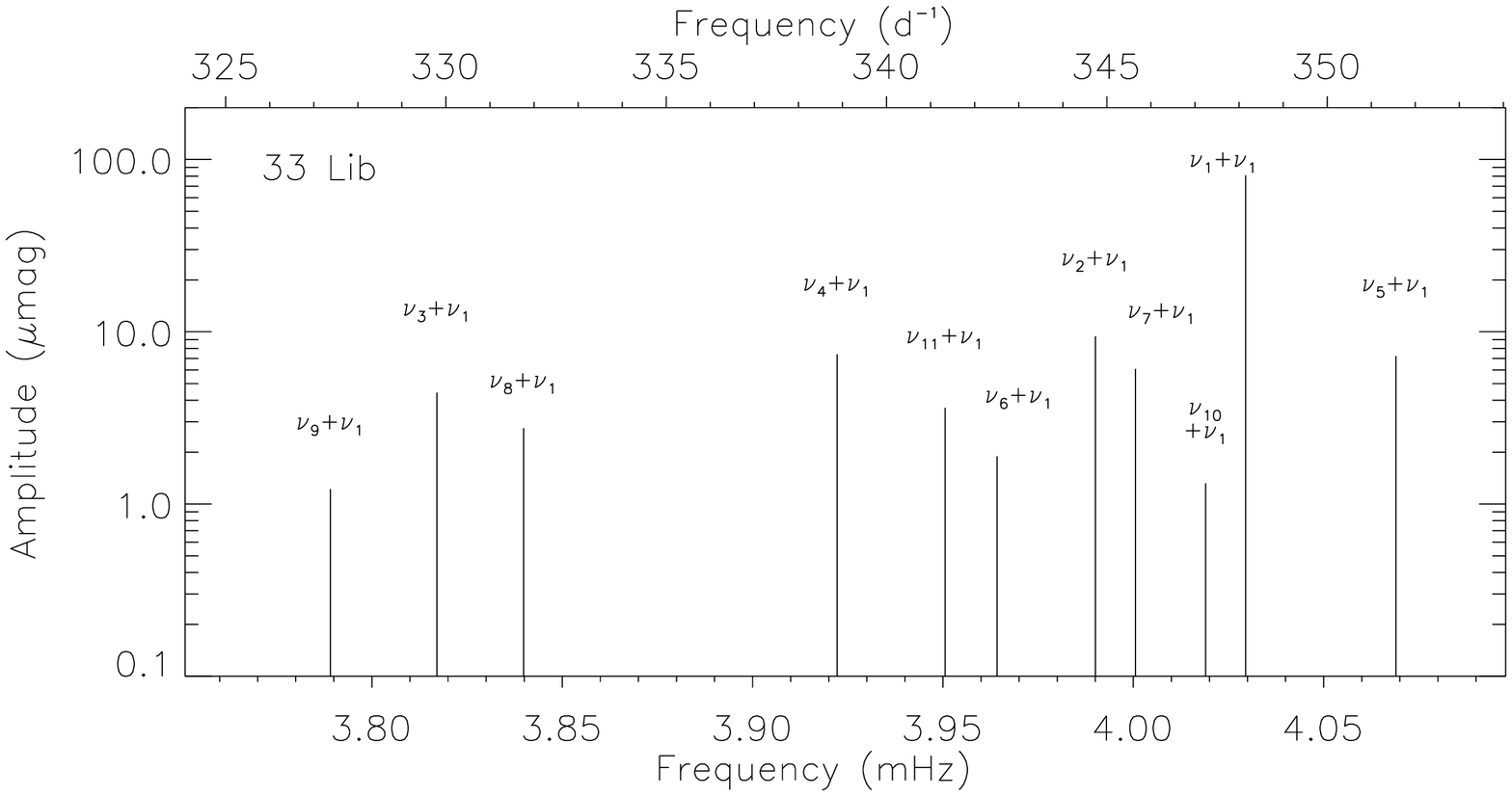}
    \end{minipage}
\end{figure}

The main findings of the K2 observations of 33\,Lib is the presence of unique non-linear interactions (Fig.\,\ref{fig:33Lib}). It is common in roAp stars to observe harmonics of the pulsation modes, since they are non-linear in nature. However, rather than a series of harmonics of the 11 modes in 33\,Lib, the authors reported the first harmonic of the principal (plus others) was accompanied by 10 peaks with frequencies of the original pulsation frequencies plus the frequency of the principal mode. This is an indication of mode coupling, i.e. frequency mixing due to non-linear effects predominately in the outer portions of the stellar envelope \citep[e.g.,][]{2014ApJ...783...89B} between the principal mode and the 10 surrounding modes. This was the first time this phenomenon was observed in an roAp star, something that would not have been possible without the \kplr\ mission.\\

\begin{figure}[htb]
    \begin{minipage}{0.25\linewidth}
    \caption{Comparison of the main pulsation frequency reported in 33\,Lib over 36\,yr. There is indication that the pulsation frequency, and thus phase, is variable over the observation period. The references for the data are: \citet{1982MNRAS.200..807K}; \citet{1991MNRAS.249..468K}; \citet{2011MNRAS.416.2669S}; \citet{2018MNRAS.480.2976H}.}
    \label{fig:33Lib_old}
    \end{minipage}
    \hspace{0.06\linewidth}
    \begin{minipage}{0.65\linewidth}
        \includegraphics[width=\linewidth]{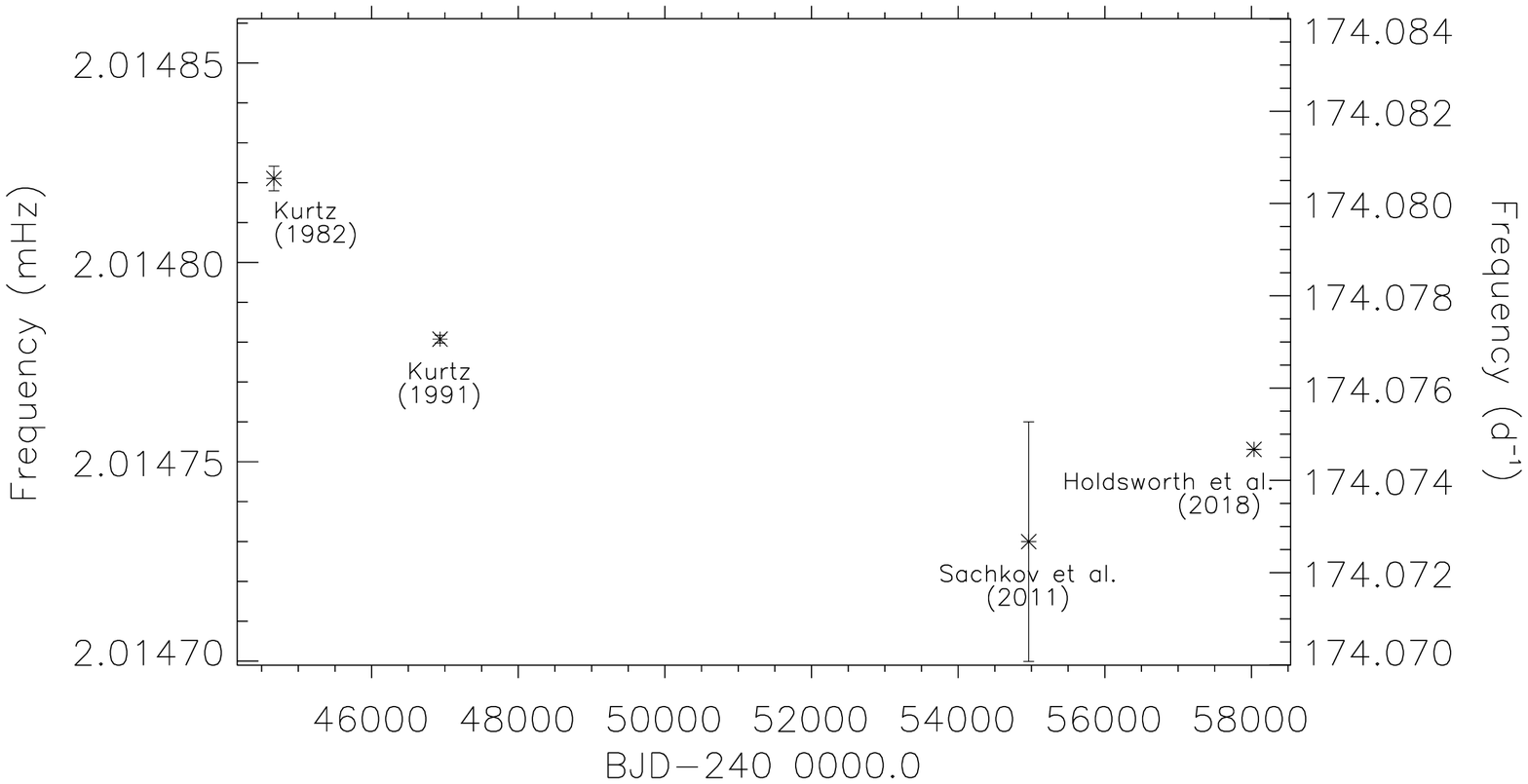}
    \end{minipage}
\end{figure}

There was no clear frequency variability in 33\,Lib during the K2 observations \citep{2018MNRAS.480.2976H}, however \citet{1991MNRAS.249..468K} showed a significant change in the pulsation frequency between observations in 1981 and 1987. The K2 observations provide a third epoch where the frequency is different from both the aforementioned data sets. Literature values of the principal pulsation frequency of 33\,Lib which cover over 36\,yr indicate significant frequency variability in this star, as shown in Fig.\,\ref{fig:33Lib_old}. However, a careful re-analysis of all available data is needed since there are further epochs of data where either an independent frequency determination and/or error measurement has not been made. These epochs have been omitted from Fig.\,\ref{fig:33Lib_old}.

\section{Summary}

Although \kplr\ did not find the `holy grail' of the roAp stars, i.e., a multi-periodic pulsator with a series of modes in the asymptotic regime, it has provided high-quality data on new and well known roAp stars, and has provided new insights into the pulsation behaviour of this class of variable star. From different pulsation axes, to potential low-frequency pulsation, binarity, non-linear interactions and significant mode distortion, \kplr\ observations have perhaps posed more questions on the fundamental understanding of these stars than they have answered.

In the cases where the full 4-yr data sets are available, it seems that all stars show a degree of frequency variability in their modes, a variability that can be different for different modes in a given star. Is this phenomenon present in all roAp stars, or indeed all pulsating stars? This has been observed before from the ground, for example in the roAp star HR\,3831 \citep{1994MNRAS.268..641K,1997MNRAS.287...69K}, although with significant gaps, but with the precision and time-space of the \kplr\ data, physical insight may now be possible. This also poses the question as to whether all roAp stars exhibit such variation, but it is only detected in the most obvious cases, or where data cover a significant time span.

It is possible that these phase variations are the first observations of stochastic perturbations of classical pulsators as discussed by \citet{2020MNRAS.492.4477A} and \citet{2020MNRAS.tmp.2755C}. In those works, the authors considered models of a damped harmonic oscillator subjected to internal or external forces, and noise, and were able to show that, given a sufficient amount of time, a random phase variation is expected. It is also expected that these random variations are different for different modes in the same star, as was discussed above in the cases of KIC\,10195926, for example. This demonstrates the continued legacy of the \kplr\ data set.

It is expected that NASA's next generation planet hunting mission, the Transiting Exoplanet Survey Satellite {\it TESS} \citep{2015JATIS...1a4003R} will revisit all of the \kplr\ observed roAp stars, and indeed all roAp stars providing a homogeneous sample to draw statistical inference from. However, with observations as short as 27-d, with a maximum of almost 1-yr of (almost) uninterrupted data, the {\it TESS} observations will not provide the precision that \kplr\ did. This, coupled with the less favourable (for roAp stars) redder passband of {\it TESS}, means that the \kplr\ data on the roAp stars will be the definitive data for precise studies of the roAp stars.\\

\section*{Conflict of Interest Statement}

The authors declare that the research was conducted in the absence of any commercial or financial relationships that could be construed as a potential conflict of interest.

\section*{Author Contributions}

The author confirms being the sole contributor of this work and has approved it for publication.

\section*{Funding}
DLH acknowledges financial support from the Science and Technology Facilities Council (STFC) via grant ST/M000877/1.

\section*{Acknowledgments}
I thank the anonymous referees for constructive comments and suggestions on the manuscript. This paper includes data collected by the Kepler mission. Funding for the Kepler mission is provided by the NASA Science Mission directorate. The author gratefully acknowledge the Kepler Science Team and all those who have contributed to making the Kepler mission possible.

\bibliographystyle{frontiersinSCNS_ENG_HUMS} % for Health, Physics and Mathematics articles
\bibliography{references}
\end{document}